\documentclass{pasj01}

\Received{}
\Accepted{}
 
\begin{document} 

\title{Three-dimensional geometry and dust/gas ratios in massive star forming regions over the entire LMC as revealed by IRSF/SIRIUS survey}

\author{Takuya \textsc{Furuta}\altaffilmark{1}$^{*}$}%
\altaffiltext{1}{Graduate School of Science, Nagoya University, Furo-cho, Chikusa-ku, Nagoya, Aichi 464-8602, Japan}
\email{t.furuta@u.phys.nagoya-u.ac.jp}

\author{Hidehiro \textsc{Kaneda}\altaffilmark{1}}%

\author{Takuma \textsc{Kokusho}\altaffilmark{1}}%

\author{Yasushi \textsc{Nakajima}\altaffilmark{2}}%
\altaffiltext{2}{National Astronomical Observatory of Japan, 2-21-1 Osawa, Mitaka, Tokyo 181-8588, Japan}

\author{Yasuo \textsc{Fukui},\altaffilmark{1}}

\author{Kisetsu \textsc{Tsuge},\altaffilmark{3}}
\altaffiltext{3}{Dr. Karl Remeis Observatory and ECAP, Universit\"{a}t Erlangen-N\"{u}rnberg, Sternwartstr. 7, 96049 Bamberg, Germany}


\maketitle

\begin{abstract}
We derive the entire dust extinction ($A_V$) map for the Large Magellanic Cloud (LMC) estimated from the color excess at near-infrared wavelengths.
Using the percentile method we recently adopted to evaluate $A_V$ distribution along the line of sight, we derive the three-dimensional (3D) $A_V$ maps of the three massive star forming regions of N44, N79 and N11 based on the IRSF/SIRIUS point source catalog.
The 3D $A_V$ maps are compared with the hydrogen column densities $N$(H) of three different velocity components where one is of the LMC disk velocity and the other two are of velocities lower than the disk velocity.
As a result, we obtain 3D dust geometry suggesting that gas collision is on-going between the different velocity components.
We also find difference in the timing of the gas collision between the massive star forming regions, which indicates that the gas collision in N44, N79 and N11 occurred later than that in 30 Doradus.
In addition, difference by a factor of two in $A_V$/$N$(H) is found between the velocity components for N44, while significant difference is not found for N79 and N11.
From the 3D geometry and $A_V$/$N$(H) in each star forming region, we suggest that the massive star formation in N44 was induced by an external trigger of tidal interaction between the LMC and the SMC, while that in N79 and N11 is likely to have been induced by internal triggers such as gas converging from the galactic spiral arm and expansion of a supershell although the possibility of tidal interaction cannot be ruled out.
\end{abstract}

\section{Introduction}
The Large Magellanic Cloud (LMC) is a dwarf galaxy with morphological features such as an off-center stellar bar and a single spiral arm, accompanied by the Magellanic Bridge which is a H\,\emissiontype{I} gas structure connecting the LMC and the Small Magellanic Cloud (SMC).
These features are considered to be evidence for the tidal interaction between the LMC and the SMC (e.g., \cite{besla_sim}; \cite{diaz_sim}).
The LMC is one of the nearest interacting galaxies at a distance of $\simeq$ 50 $\rm kpc$ (\cite{proximity}) with almost face-on orientation ($i\sim35^{\circ}$; \cite{inclination}), which enables us to investigate the spatial distribution of the interstellar medium (ISM) with high-angular resolution ($<$1 pc) in almost two dimensions.
In addition, the metallicity of the LMC ($Z \sim$ 0.5 $\rm Z_{\odot}$; \cite{metal}) is similar to a value typical of the ISM at redshift $z\sim 0.5$ (\cite{redshift}) corresponding to a relatively active phase of the star formation history in the Universe.
Furthermore, active star formation is on-going in the LMC, where 30 Doradus (30 Dor) is the most active starburst region in the Local Group (e.g., \cite{kennicutt}).
Thanks to these characteristics, the LMC is a site suitable for investigating how the galactic interaction is taking place and how massive star formation is induced in the low metallicity environments.

Recent studies using H\,\emissiontype{I} and CO data suggest that the massive star formations in 30 Dor and N159 in the LMC H\,\emissiontype{I} ridge region were induced by the galactic interaction between the LMC and the SMC (\cite{fukui_2017}, \yearcite{n159_fukui}; \cite{n159_tokuda}).
\citet{fukui_2017} identify two velocity H\,\emissiontype{I} components; one is the velocity component of the LMC disk and the other is a low velocity component relative to the disk velocity.
\citet{fukui_2017} and \citet{tsuge} identify an intermediate velocity component between them for 30 Dor and N159, which is considered to be evidence for gas collision.
In addition, they evaluate the dust/gas ratio from the comparison of the H\,\emissiontype{I} intensity with the dust optical depth, and find that the dust/gas ratio for the LMC H\,\emissiontype{I} ridge region is lower than that for the LMC stellar bar region.
Based on the low dust/gas ratio and numerical simulations (e.g., \cite{bekkia}, \yearcite{shock_sim}; \cite{yozin}), they suggest that low-metallicity gas from the SMC ($Z \sim$ 0.2 $\rm Z_{\odot}$; \cite{russell}) is falling onto the LMC disk.
More recently, \citet{tsuge_2021} investigate the spatial correlation between the intermediate velocity component and massive stars in the entire LMC field, and suggest that the galactic interaction may have induced a large part ($\sim$70\%) of the massive star formation in the LMC.
In order to investigate such gas collisions between the gas in the LMC and an inflow gas from the SMC, it is important to study the dust/gas ratio and three-dimensional (3D) dust geometry over the whole LMC.

\citet{furuta} constructed the near-infrared (NIR) dust extinction ($A_V$) map of the LMC H\,\emissiontype{I} ridge region to evaluate the dust/gas ratio.
They proposed a method to decompose $A_V$ into different velocity components from the comparison of $A_V$ with hydrogen column densities of different velocities.
They decomposed $A_V$ into the three velocity components mentioned above, and found that the dust/gas ratio of the low velocity component is significantly lower than that of the disk velocity one.
Furthermore, \citet{furuta_2021} developed a new method to evaluate 3D dust geometry from the NIR dust extinction, and applied it to the H\,\emissiontype{I} ridge region.
They decomposed the resultant 3D $A_V$ map into the three velocity components using the fitting method described in \citet{furuta}, and obtained the detailed dust geometry of the different velocity components, from which they found that the gas collision with the inflow gas from the SMC had occurred in 30 Dor prior to N159.

In the present paper, we extend the analysis of the LMC H\,\emissiontype{I} ridge region by \citet{furuta_2021} to the entire LMC field.
We especially focus on the massive star forming regions of N44, N79 and N11 for which the tidal interaction between the Magellanic Clouds are reported (\cite{tsuge}, \yearcite{tsuge_2021}).
This paper is organized as follows: we summarize the data in section 2, and give a brief review of the method to evaluate the 3D dust extinction in section 3.
The resultant dust extinction maps and comparison with gas observations are presented in section 4.
We discuss the 3D geometry and origin of massive star forming regions in section 5, and summarize our results in section 6.

\section{The data}
\subsection{Point source catalog observed with IRSF/SIRIUS}\label{sec:sample}
We used the NIR ($J$, $H$ and $K_{S}$ bands) Magellanic Clouds point source catalog observed with the InfraRed Survey Facility (IRSF) 1.4 m telescope (\cite{irsf}; hearafter the IRSF catalog).
The IRSF catalog covers 40 $\rm{deg}^2$ area of the LMC.
The 10$\sigma$ limiting magnitudes of the $J$, $H$ and $K_{S}$ bands are 18.8, 17.8 and 16.6 mag, respectively, which are three magnitudes deeper than those of the 2MASS catalog.
In the IRSF catalog, systematic photometric errors due to inappropriate flat-field correction were reported, and thus \citet{furuta} updated the IRSF catalog by applying the corrected flat-field images to the raw data of the catalog ( see section 2 of \cite{furuta}).
In this study, data is selected from the updated IRSF catalog that satisfies the following criteria: (1) the $J$, $H$ and $K_{S}$ bands magnitudes are brighter than the $10\sigma$ limiting magnitudes of the original IRSF catalog, (2) the shapes of sources are ``point like'' (``quality flag'' is 1) in all the bands and (3) at least 8 dithered frames are combined in all the bands. 

\begin{figure*}
\begin{center}
  \includegraphics[width=16cm]{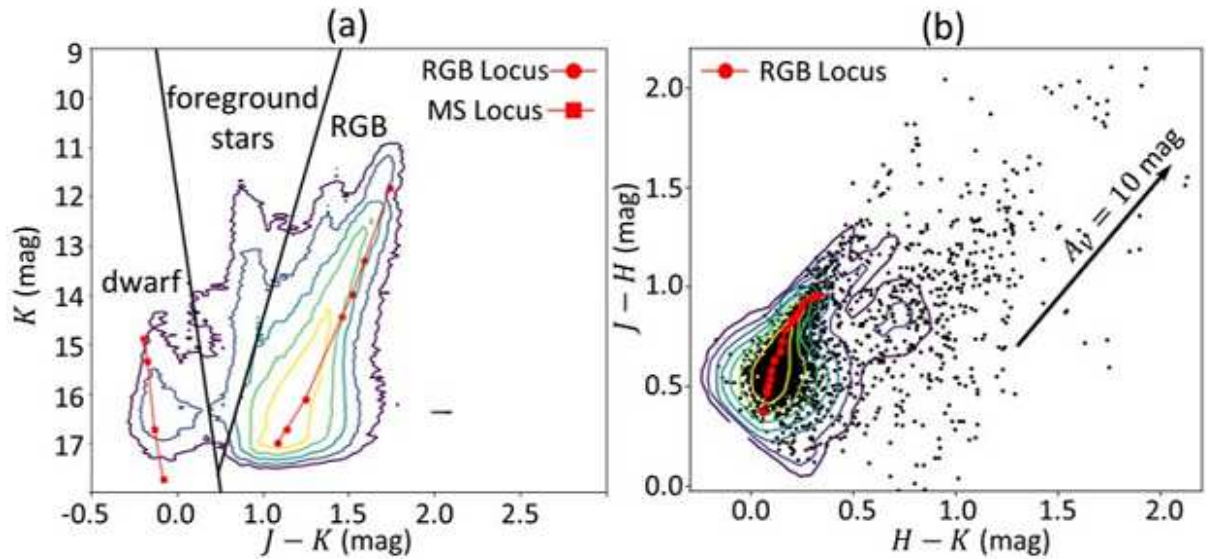} 
 \end{center}
\caption{(a) Color-magnitude ($J-K$ and $K$) diagram of the samples of the entire LMC. Contours show the number densities binned by 0.05 mag. The contour levels are $10^{1.5}$--$10^{3.5}$ with a step of $10^{0.5}$. Black solid lines are boundary positions to classify stellar populations into three categories of dwarf, foreground stars and RGB stars. The lines connected with the circles and squares are the loci of RGB and main sequence stars, respectively, derived from \citet{allen}. (b) Color-color ($H-K$ and $J-H$) diagram of the RGB samples selected from panel (a). Contours are same as those in panel (a) but the levels are $10^{1.5}$--$10^{4.5}$ with a step of $10^{0.5}$. The line connected with the red circles is intrinsic colors of RGB stars derived from \citet{tr_giant}. Black arrow is the reddening vector corresponding to $A_V=10$ mag. Black points are YSO candidates matched to the RGB samples.}\label{fig:color_mag}
\end{figure*}

\begin{figure}
\begin{center}
  \includegraphics[width=8cm]{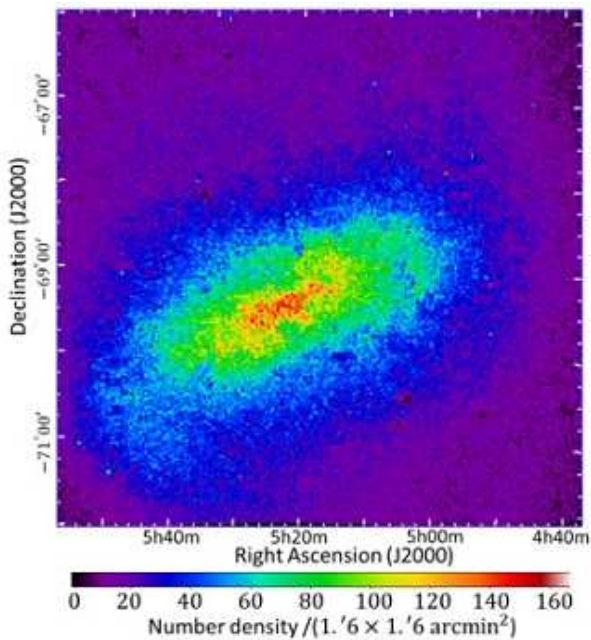} 
 \end{center}
\caption{RGB star number density map of the LMC. The grid size is \timeform{1.'6}$\times$\timeform{1.'6}. The color scale shows the number of stars included in a grid.}\label{fig:number_density}
\end{figure}

Since Galactic foreground stars leading to under-estimation of the dust extinction are included in the IRSF catalog, we identify stellar populations using the classification by \citet{furuta}.
Stellar populations are classified into three categories which primarily consist of Galactic foreground contamination, red giant branch (RGB) stars and dwarf stars, using a color-magnitude ($J-K$ vs. $K$) diagram (figure \ref{fig:color_mag}a).
In the present paper, only the objects classified as RGB stars are used, the number density map of which is shown in figure \ref{fig:number_density}.

In principle, intrinsically red stars such as Young Stellar Objects (YSOs) can contaminate the selected RGB samples.
To check the contamination of YSOs, we match the RGB samples with the YSO candidates in the whole LMC derived from \citet{whi08}, \citet{gru09} and \citet{sea14}.
We also use the list of YSOs presented by \citet{car12}, who identify YSOs in the nine star forming regions in the LMC.
The total number of YSO candidates is 5714.
A matching distance of \timeform{5''} is adopted, which is the same value as used by \citet{sea14}.
As a result of the matching, only $\sim 0.13\%$ (2033/1522279) of the RGB samples in the whole LMC are identified as YSOs.
The YSOs matched to our RGB samples are plotted as black points on a color-color ($H-K$ and $J-H$) diagram in figure \ref{fig:color_mag}b.
The fraction of contamination of YSOs increases with the dust extinction ($A_V$); the fraction is $0.31\%$ (877/278612) for samples with $A_V> 1.0$ mag, while that is $1.12\%$ (549/49008) for samples with $A_V>2.0$ mag.
In this way, a small fraction of YSOs contaminates the RGB samples, leading to over-estimation of $A_V$ especially in star forming regions.
We eliminate such intrinsically red stars from the RGB samples using the percentile method in deriving $A_V$ maps, as described in section \ref{sec:method_av}.
\subsection{Gas tracer}

For comparison of the dust extinction with the hydrogen column density, we use the H\,\emissiontype{I} data observed with the Australia Telescope Compact Array (ATCA) and Parkes (\cite{himap}), and the $^{12}{\rm CO}\ (J=$1--0) data observed with the NANTEN 4 $\rm m$ telescope (\cite{comap}).
\citet{fukui_2017} and \citet{tsuge} calculate the velocity relative to the galactic rotation defined as $V_{\rm offset}$, and decompose the H\,\emissiontype{I} and CO integrated intensity maps into three velocity components defined as the L- (named after the first letter of ``Low''), I- (``Intermediate'') and D- (``Disk'') component according to the integrated velocity range of $V_{\rm offset}$=$-$100 to $-$30, $-$30 to $-$10 and $-$10 to 10 $\rm km\ s^{-1}$, respectively (for the analysis and an example of the H\,\emissiontype{I} spectra, see \cite{tsuge}).

Following \citet{tsuge}, the H\,\emissiontype{I} intensity maps are converted into H\,\emissiontype{I} column density ($N({\rm H\,\emissiontype{I}})$) maps using the conversion factor, $X_{\rm H\,\emissiontype{I}}=1.82 \times 10^{18}\ \rm{cm^{-2}}/({\rm K\ km\ s^{-1}})$ (\cite{nh_conv}), while the CO intensity maps are converted into the $\rm H_{2}$ column density ($N({\rm H_{2}})$) maps using the CO to $\rm H_{2}$ conversion factor, $X_{\rm CO}=7\times10^{20}\ {\rm cm^{-2}}/(\rm{K\ km\ s^{-1}})$ which is the averaged $X_{\rm CO}$ over the LMC (\cite{h2_conv}). 
The total hydrogen column densities $N$(H) are calculated for the L-, I- and D-components as
\begin{equation}
N({\rm H})\ =\ N({\rm H\,\emissiontype{I}})\ + \ 2N({\rm H_{2}}).
\label{nh}
\end{equation}
The spatial resolutions of the gas maps are reduced to be the same as that of the dust extinction map by convolving them with a boxcar kernel of \timeform{5'}$\times$\timeform{5'}.

\subsection{Dust emission}
To compare the dust extinction with the dust emission, we use the dust optical depth at 353 GHz ($\tau_{353}$) derived from the Planck and IRAS data (\cite{tau_map}).
The comparison between the dust extinction and emission gives an insight into the dust geometry because the dust extinction only traces the dust located in front of stars, while the dust emission traces all the dust existing along the line of sight.
As a dust emission map of the LMC, we use the $\tau_{353}$ map created by \citet{tsuge}, who subtract the Galactic foreground $\tau_{353}$ values.
The spatial resolution of the $\tau_{353}$ map is adjusted to that of the dust extinction map by the same procedure as the gas maps.

\section{Method}
\citet{furuta_2021} developed a new method to evaluate the 3D dust geometry for study of the LMC H\,\emissiontype{I} ridge region.
The detailed procedure is described in their paper (see section 3 of \cite{furuta_2021}).
Here, we introduce a brief review of the method.
\subsection{Derivation of dust distribution along the line of sight}\label{sec:method_av}
\begin{figure*}
 \begin{center}
  \includegraphics[width=16cm]{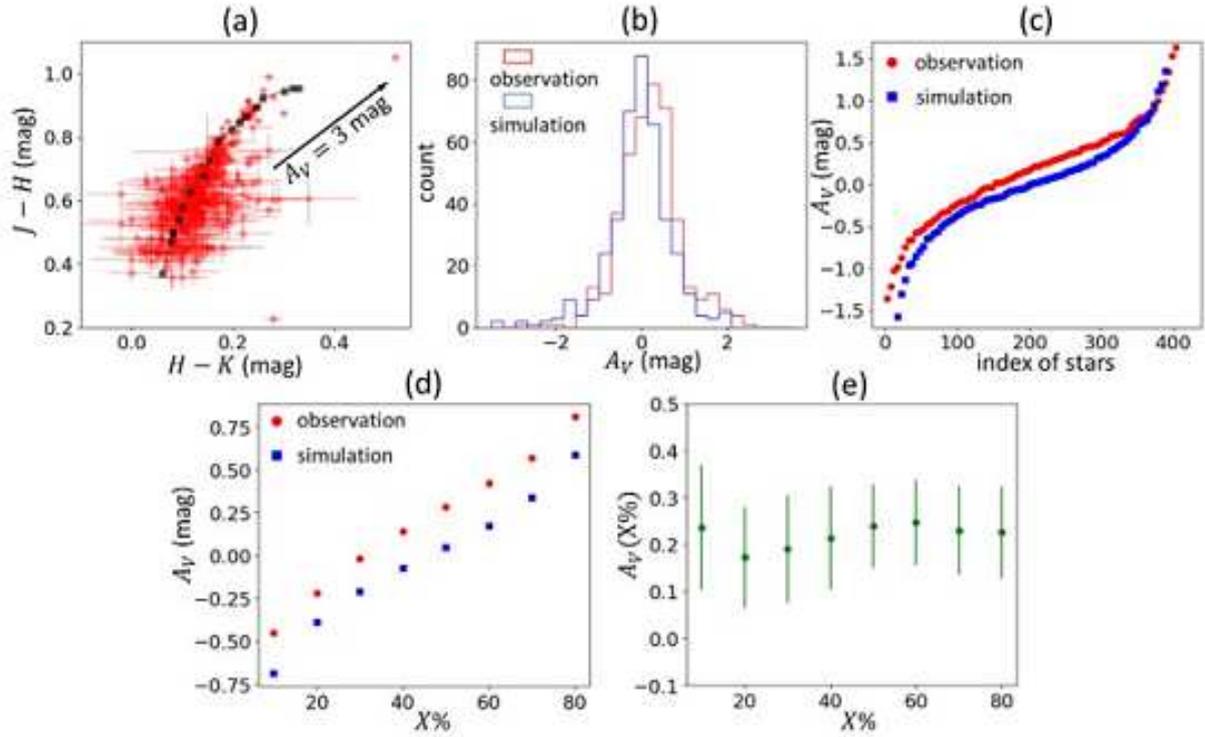} 
 \end{center}
\caption{Procedure to estimate $A_V$ along the line of sight in a spatial bin of \timeform{5'} for a low H\,\emissiontype{I} column density region. (a) Example of the color-color ($H-K$ and $J-H$) diagram of stars in a spatial bin centered at $(\alpha, \delta)_{\rm J2000.0} =$ (\timeform{5h32m}, $-$\timeform{69D14'}). Black squares and red circles are intrinsic and observed colors of RGB stars, respectively. The reddening vector is shown as black arrow corresponding to $A_V$=3 mag.
(b) Histograms of observed (red) and simulated (blue) $A_V$ for stars in panel (a).
(c) Observed (red) and simulated (blue) $A_V$ distributions plotted at every 5 stars as a function of the cumulative number of stars for samples in panel (a).
(d) Mean $A_V$ included in $X\%$ to $(X+10)\%$ percentile of $A_V$ shown in panel 
(e) Difference in $A_V$ at each $X\%$ between observed and simulated values shown in panel (d).}\label{fig:procedure}
\end{figure*}

We below show the calculation procedure of dust distribution along the line of sight for a low H\,\emissiontype{I} column density region ($<5\times10^{20}\  \rm{cm^{-2}}$) at $(\alpha, \delta)_{\rm J2000.0} =$ (\timeform{5h32m}, $-$\timeform{69D14'}) where we expect to obtain the distribution in the case of no appreciable dust.
In figure \ref{fig:procedure}a, we present the color-color diagram (CC diagram) in the spatial bin of \timeform{5'}$\times$\timeform{5'} centered at the region.
We first estimate $A_V$ for individual stars from the color excess between the observed and intrinsic colors of RGB stars along the reddening vector (black arrow in figure \ref{fig:procedure}a) on the CC diagram.
Here, we assume the intrinsic colors derived from \citet{tr_giant}, and reddening vector of \citet{red_law2}.

Here, it is noted that recent studies show the dependence of RGB colors on stellar metallicity (e.g., \cite{val04}) together with the metallicity gradient in the LMC (e.g., \cite{cho21}), which can lead to an unintended bias in $A_V$ at different galactocentric distances across the LMC with the intrinsic colors of \citet{tr_giant} that do not consider the metallicity dependence. 
Therefore, we check how much the intrinsic colors are affected by the metallicity gradient.
\citet{cho21} find the average metallicity of [Fe/H]$=-0.42$ dex and a shallow metallicity gradient of $-0.008$ dex $\rm kpc^{-1}$ from the galactic center in the LMC using the CM diagram of RGB stars.
Our $A_V$ map covers radii up to 2.6 kpc from the center, and thus the stellar metallicity can vary by at most $-0.02$ dex.
By changing the stellar metallicity from $-0.42$ to $-0.44$ dex considering the above metallicity gradient, the M0-type RGB colors that depend on stellar metallicity proposed by \citet{val04} change from 0.6922 to 0.6892 mag for $J-H$ and from 0.1059 to 0.1055 mag for $H-K$.
The changes in the colors are much smaller than the average photometric errors of 0.03 and 0.06 mag for the $J-H$ and $H-K$ colors of our samples, respectively, and thus the metallicity gradient in the LMC would not change our results significantly.

As a second step, we sort the estimated $A_V$ from low to high for each spatial bin.
The sorted $A_V$ distribution is shown by the red histogram in figure \ref{fig:procedure}b, while the sorted $A_V$ values at every 5 stars are shown by the red circles as a function of the cumulative number of stars in figure \ref{fig:procedure}c.
The $A_V$ values in figure \ref{fig:procedure}c monotonically increase due to the dust existing along the line of sight.
However, even when no dust exists along the line of sight, the sorted $A_V$ values increase due to the $A_V$ scatter caused by the photometric errors.
Thus, as a third step, we simulate the $A_V$ scatter by a Monte Carlo simulation to evaluate this effect.

We generate a random set of the $J-H$ and $H-K$ colors for individual stars on the CC diagram in each spatial bin under the assumption that the stellar colors follow a 2D Gaussian distribution.
Here, we adopt the observed colors and photometric errors as the center and the standard deviation of the Gaussian distribution, respectively.
$A_V$ for each simulated star is calculated by the same procedure as the first step.
The blue histogram and squares in figures \ref{fig:procedure}b and \ref{fig:procedure}c, respectively, show the resultant distribution of the simulated values, which is considered to be $A_V$ scatter due to the photometric errors alone.

As a next step, the mean percentile values for both observed and simulated $A_V$ are calculated in the range of $X$\% to $(X+10)$\% percentile for $X=$10\% to 80\% with a step of 10\%.
Here, we adopt not $100\%$ but $90\%$ as the upper limit of percentile for calculation of $A_V$ to avoid over-estimation of $A_V$ due to the contamination of intrinsically red stars such as YSOs and Asymptotic Giant Branch (AGB) stars as described in \citet{dobashi}.
Since the fraction of contamination of YSOs is small ($\sim 1\%$) even in high $A_V$ regions as mentioned in section \ref{sec:sample}, the intrinsically red stars should be eliminated from our samples through the percentile method.
Red circles and blue squares in figure \ref{fig:procedure}d are the resultant mean observed and simulated percentile $A_V$ for each $X$\%, respectively.
In order to evaluate the intrinsic $A_V$ distribution, we finally subtract the simulated percentile $A_V$ values from the observed one at each percentile range (figure \ref{fig:procedure}e), which is defined as $A_V$($X$\%).
For example, $A_V$(80\%) corresponding to the near-total integrated dust extinction is the mean $A_V$ from 80\% to 90\% percentile of observed $A_V$ after subtracting simulated percentile $A_V$ in the same percentile range.

The uncertainty of $A_V$($X$\%) is measured by the error propagation, $\delta A_V(X\%)=\sqrt{\sigma_{A_{V,X}}^2 + \sigma_{{d},X}^2}$, where $\sigma_{A_{V,X}}$ is the uncertainty of percentile $A_V$, while $\sigma_{{d},X}$ virtually corresponds to the uncertainty of the distance.
$\sigma_{A_{V,X}}$ is calculated by propagation of the $A_V$ uncertainty included in $X$\% to $(X+10)$\% percentile.
On the other hand, $\sigma_{{d},X}$ is calculated based on the simple count uncertainty (i.e., $\sqrt{N}$; $N$ is the number of the total stars contained in the bin), and determined by half the difference in $A_V$ between the index of the integer part of ($i_{\rm{center}}+\sqrt{N}$) and that of ($i_{\rm{center}}-\sqrt{N}$), where $i_{\rm center}$ is the median index of stars in the range of $X$\% to $(X+10)$\% percentile.
The uncertainty of $\delta A_V(X\%)$ is used for the comparison of the dust extinction with the hydrogen column density to evaluate the dust geometry and the dust/gas ratio (subsection \ref{sec:method_avnh}).

$A_V$($X$\%) in figure \ref{fig:procedure}e is almost constant for all $X$\%, which indicates that dust clouds associated with the LMC are not detected significantly at this spatial bin, as expected from the low column density region.
The constant $A_V$ of 0.2 mag is likely to be caused by the Galactic foreground extinction and is actually consistent with the foreground extinction $\sim 0.2$ mag (\cite{dobashi}).

\begin{figure*}
 \begin{center}
  \includegraphics[width=16cm]{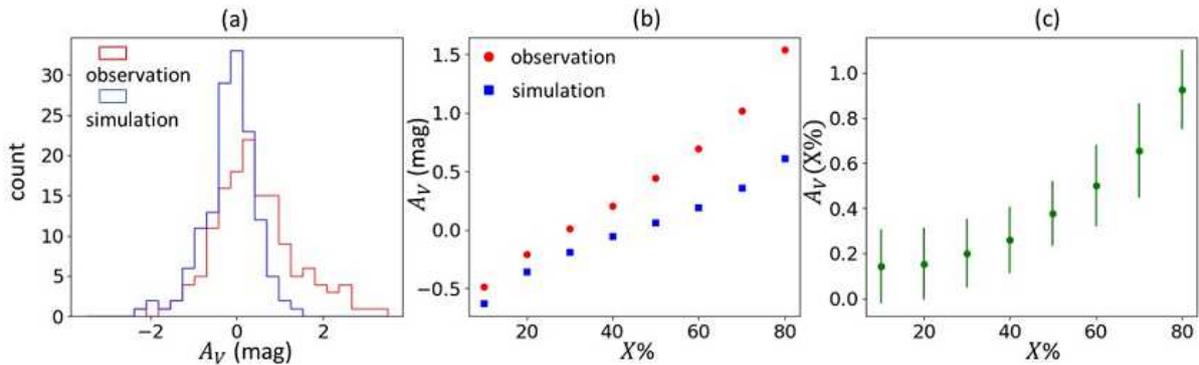} 
 \end{center}
\caption{Same diagram as in panels (b), (d) and (e) of figure \ref{fig:procedure} but for the region having a high column density of the D-component at $(\alpha, \delta)_{\rm J2000.0} =$ (\timeform{5h23m}, $-$\timeform{68D03'}) in N44.}\label{fig:eg_los}
\end{figure*}

In figure \ref{fig:eg_los}, we present another result for the spatial bin centered at $(\alpha, \delta)_{\rm J2000.0} =$ (\timeform{5h23m}, $-$\timeform{68D03'}) in N44 where the D-component is detected (the position is plotted in figure \ref{fig:n44_av_w_gas}).
The observed histogram in figure \ref{fig:eg_los}a has a peak around $A_V=0$ mag and shows $A_V$ systematically larger ($>1.5$ mag)  than the simulated histogram.
$A_V(X\%)$ from $X$=10\% to 30\% in figure\ref{fig:eg_los}c is almost constant value of $\sim 0.2$ mag corresponding to the Galactic foreground extinction.
On the other hand, $A_V(X\%)$ rapidly increases at $X=40\%$ and onward.
These results indicate that a dust cloud is present from $X=40\%$ to 80\%, which is consistent with the existence of the high column density gas of the D-component.
\subsection{Decomposition of dust extinction into different velocities}\label{sec:method_avnh}
We describe the procedure to decompose $A_V$ into different velocity components as proposed by \citet{furuta}.
The decomposition is performed from the comparison of the dust extinction map with the hydrogen column density maps of different velocities on a pixel by pixel basis.
The fitting method is based on $\chi^2$ minimization using $\delta A_V(X\%)$ calculated in subsection \ref{sec:method_av}.
The fitting function is the linear regression described as
\begin{eqnarray}\label{eq:regress1}
A_V=\sum_k a_k {[N({\rm H}\,\emissiontype{I})_{k}+2x_{k} W({\rm CO})_{k}]} \; \; \; \; \; \; (k=\rm L,\ I\ and\ D), 
\end{eqnarray}
where $a_k$ and $x_k$ are free parameters corresponding to the dust/gas ratios (i.e., $A_V$/$N$(H)) and $X_{\rm CO}$ of the $k$ ($k=$L, I and D) components, respectively.
$N({\rm H}\,\emissiontype{I})_{k}$ and $ W({\rm CO})_{k}$ are the ${\rm H}\,\emissiontype{I}$ column density and CO intensity of the $k$ component, respectively.
Applying this fitting to each $A_V$($X \%$) map from $X=10$\% to $80$\%, we  decompose the $A_V$ distribution along the line of sight into the three different velocity components.
From the changes of the resultant $a_k$ (i.e., decomposed $A_V$ of the $k$-component) with $X$\%, we evaluate the 3D dust geometry of different velocities (see figures 13 and 14 of \cite{furuta_2021}).

\section{Result}
\subsection{Cumulative dust extinction map over the LMC}\label{sec:res_entire}
\begin{figure*}
\begin{center}
\includegraphics[width=16cm]{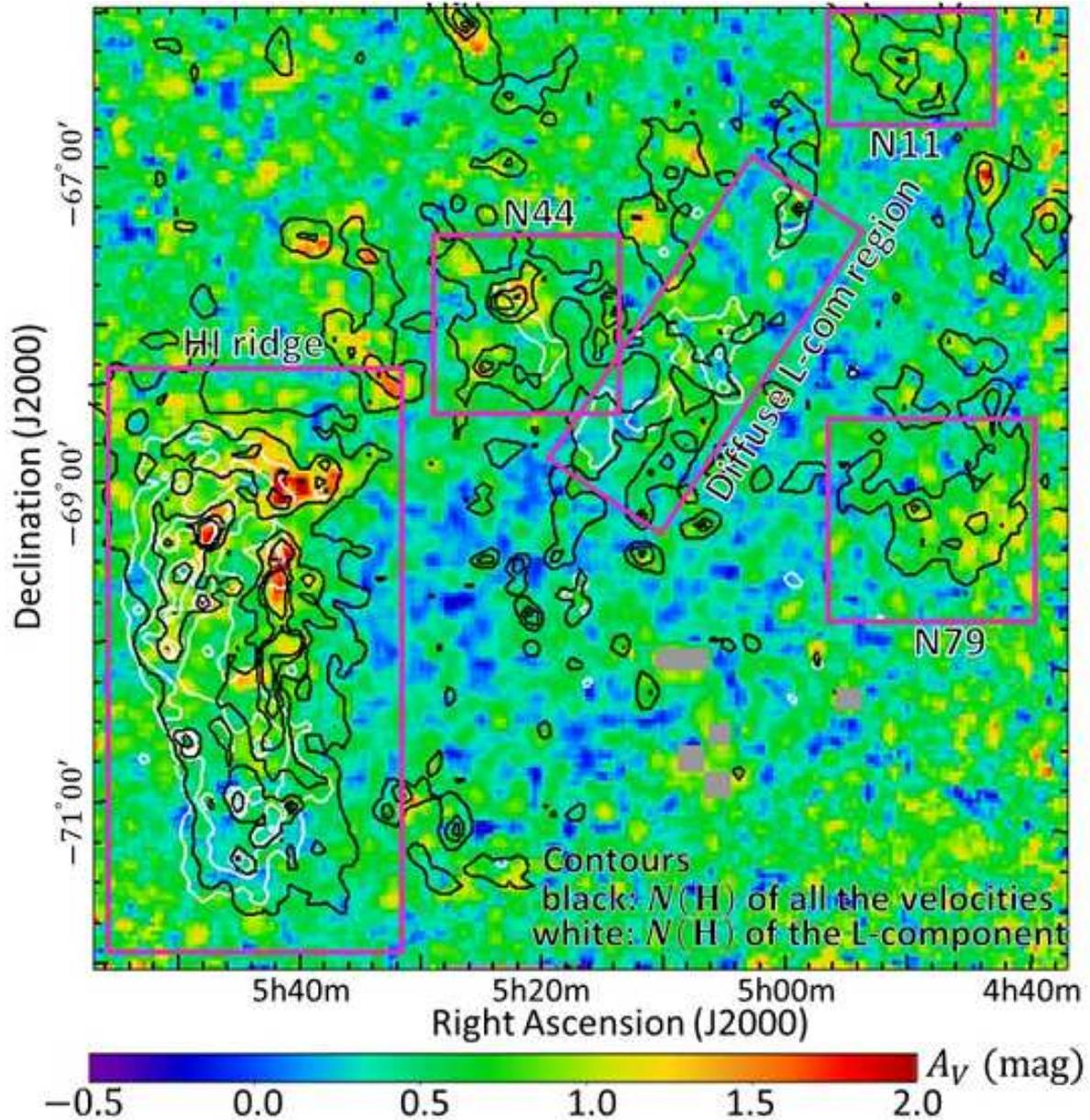} 
\end{center}
\caption{Total integrated dust extinction map ($A_V(80\%)$) in the entire LMC field. The spatial resolution is \timeform{5'} with a grid size of \timeform{1.'6}. Black contours show the total hydrogen column density covering all the velocity ranges of the L-, I- and D-components with the contour levels of ($2.0$--$8.0$)$\times 10^{21}$ $\rm cm^{-2}$ in steps of $2.0\times10^{21}$ $\rm cm^{-2}$. White contours show the total hydrogen column density of the L-component with the levels of ($0.7,\ 1.6,\ 2.5,\ 3.4,\ 4.2,\ 5.1$ and $6.0$)$\times 10^{21}$ $\rm cm^{-2}$. Gray regions are masked bins due to the inappropriate observational data.}\label{fig:av_allLMC}
\end{figure*}

\begin{figure}
 \begin{center}
  \includegraphics[width=8cm]{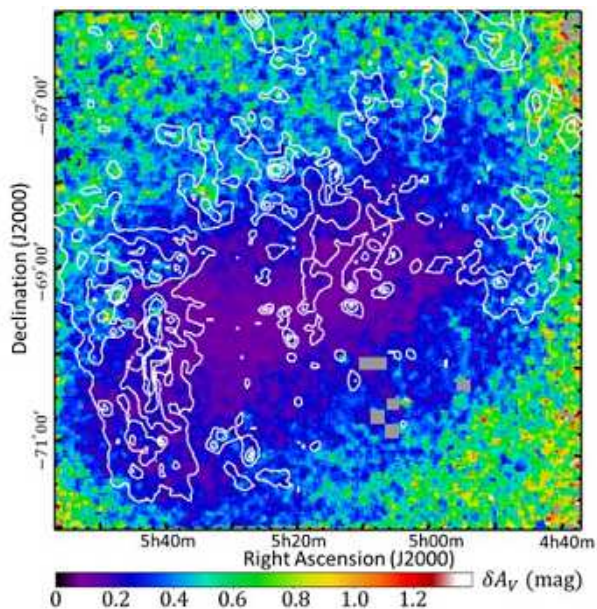} 
 \end{center}
\caption{Uncertainty map ($\delta A_V(80\%)$) of the total integrated dust extinction map in figure \ref{fig:av_allLMC}.
White contours are the same as the black contours in figure \ref{fig:av_allLMC}.}\label{fig:averr_allLMC}
\end{figure}

In figure \ref{fig:av_allLMC}, we show the $A_V$(80\%) map (i.e., near-total integrated dust extinction map) of the entire LMC field.
The spatial resolution of the map is \timeform{5'} with a grid size of \timeform{1.'6}.
The uncertainty of the map ($\delta A_V (80\%)$) is shown in figure \ref{fig:averr_allLMC}.
The uncertainty is relatively large near the edge of the map due to low number densities of stars.
In these maps, we mask the regions where we see failure in the read-out of the data or inappropriate zero-point magnitude calibration.

\begin{figure}
 \begin{center}
  \includegraphics[width=8cm]{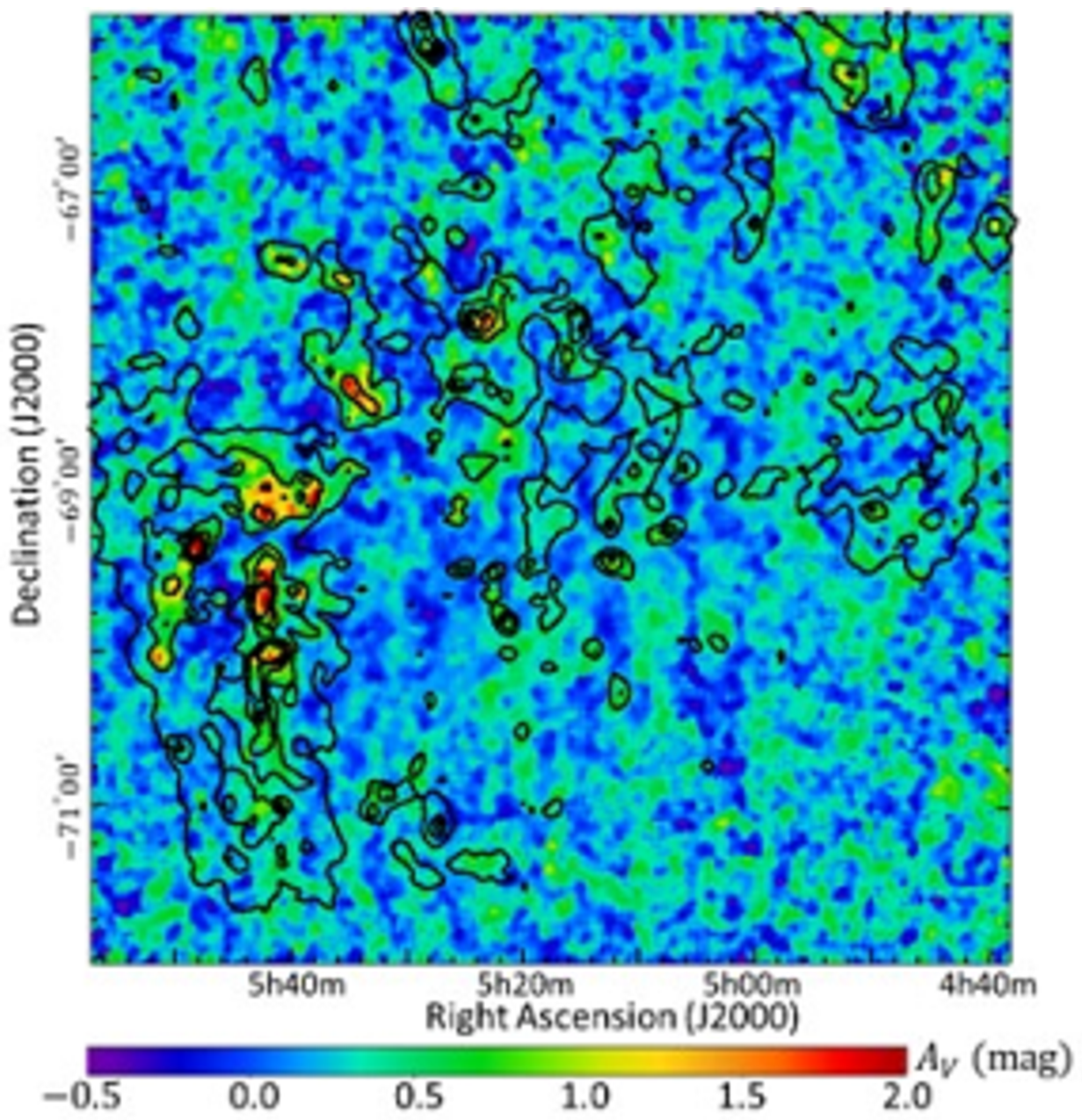} 
 \end{center}
\caption{Dust extinction map in the entire LMC field using the 2MASS catalog (\cite{dobashi}). The area, the color scale and black contours are same as those in figure \ref{fig:av_allLMC}.}\label{fig:2mass_av}
\end{figure}

We compare the $A_V$(80\%) map with the dust extinction map created by \citet{dobashi} using the 2MASS catalog in figure \ref{fig:2mass_av}.
The spatial distribution of high $A_V$ regions such as the H\,\emissiontype{I} ridge region and N44 is consistent between the $A_V$ maps in figures \ref{fig:av_allLMC} and \ref{fig:2mass_av}.
The $A_V$(80\%) map significantly detects dust clouds in N79 and around $(\alpha, \delta)_{\rm J2000.0} =$ (\timeform{5h25m}, $-$\timeform{66D15'}), which are not detected in the map of \citet{dobashi}.
This is probably owing to the fact that the limiting magnitude of the IRSF catalog is three magnitude deeper than that of the 2MASS catalog.
We compare the $A_V$ map of this study and that of \citet{dobashi} on a pixel by pixel basis and find a small but systematic difference in $A_V$ ($0.3$ mag), which can be seen especially in low hydrogen column density regions in figures \ref{fig:av_allLMC} and \ref{fig:2mass_av}.
This is probably due to the difference in the method to estimate $A_V$; we select RGB stars and use the several intrinsic colors based on the spectral types of RGB stars, whereas \citet{dobashi} did not classify stellar populations and used an intrinsic color averaged for the stars of any spectral types and stellar populations in their reference field (i.e., extinction-free region).
In the latter case, the $A_V$ values can vary depending on the spectral types and populations of the stars included in each observed field, and thus it is difficult to detect diffuse dust clouds.
On the other hand, our dust extinction map is considered to detect diffuse dust clouds because the dust/gas ratio ($A_V/N$(H)) of the D-component in our study is consistent with that of the LMC estimated from the previous studies as discussed in section \ref{sec:dis_avnh}.
 
In figure \ref{fig:av_allLMC}, we recognize that the dust extinction correlates well with the total hydrogen column density.
Relatively high $A_V$ regions ($A_V > 1.0$ mag) can be seen in the massive star forming regions of H\,\emissiontype{I} ridge, N44, N79 and N11.
On the other hand, in the region where \citet{tsuge_2021} identify the diffuse L-component (hereafter the Diffuse L-com region), no dust cloud is detected.
%
%
%
%
%
%
\subsection{Cumulative dust extinction toward individual star forming regions}
\begin{figure*}
 \begin{center}
  \includegraphics[width=15cm]{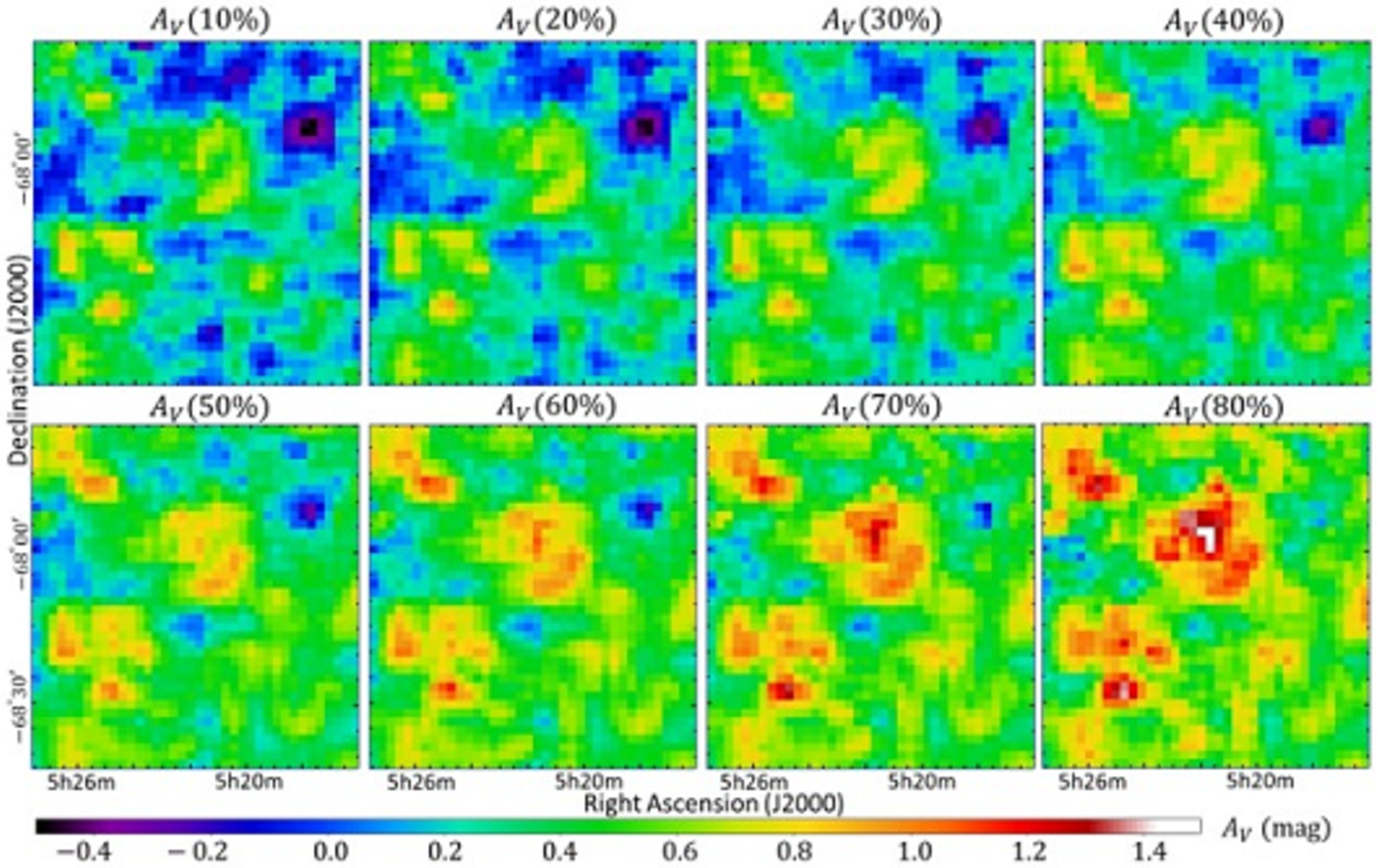} 
 \end{center}
\caption{Integrated dust extinction maps of N44 from observers to stars included in $X\%$ percentile for $X=10\%$ to $80\%$ in steps of $10\%$.}\label{fig:all_n44}
\end{figure*}

\begin{figure*}
\begin{center}
\includegraphics[width=16cm]{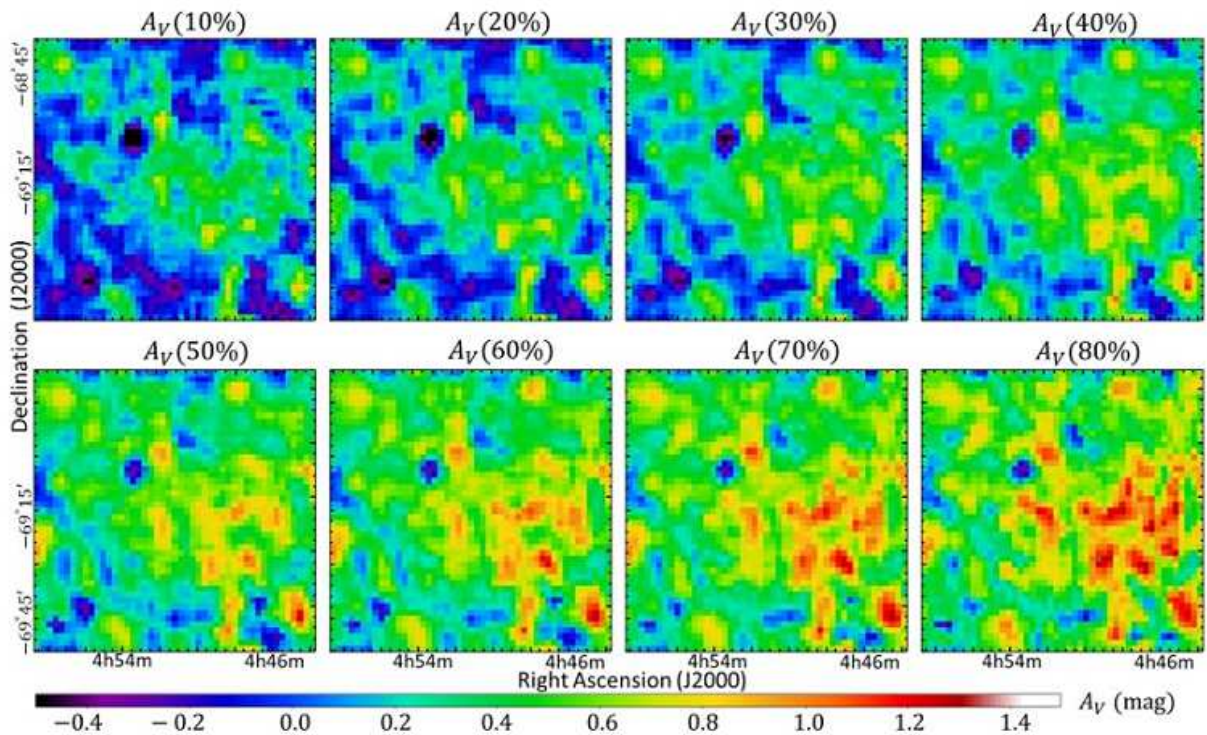} 
\end{center}
\caption{Same maps as figure \ref{fig:all_n44} but for N79.}\label{fig:percentile_n79}
\end{figure*}

\begin{figure*}
 \begin{center}
  \includegraphics[width=14cm]{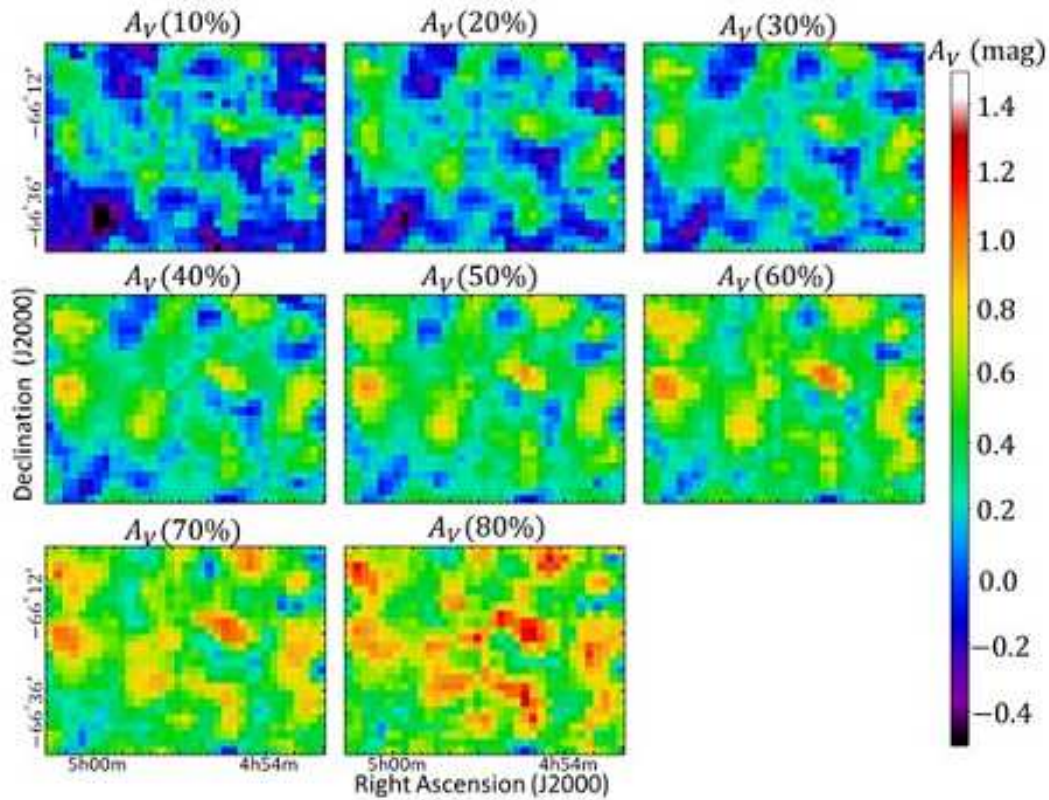} 
 \end{center}
\caption{Same maps as figure \ref{fig:all_n44} but for N11}\label{fig:percentile_n11}
\end{figure*}

\begin{figure*}
  \begin{center}
  \includegraphics[width=14cm]{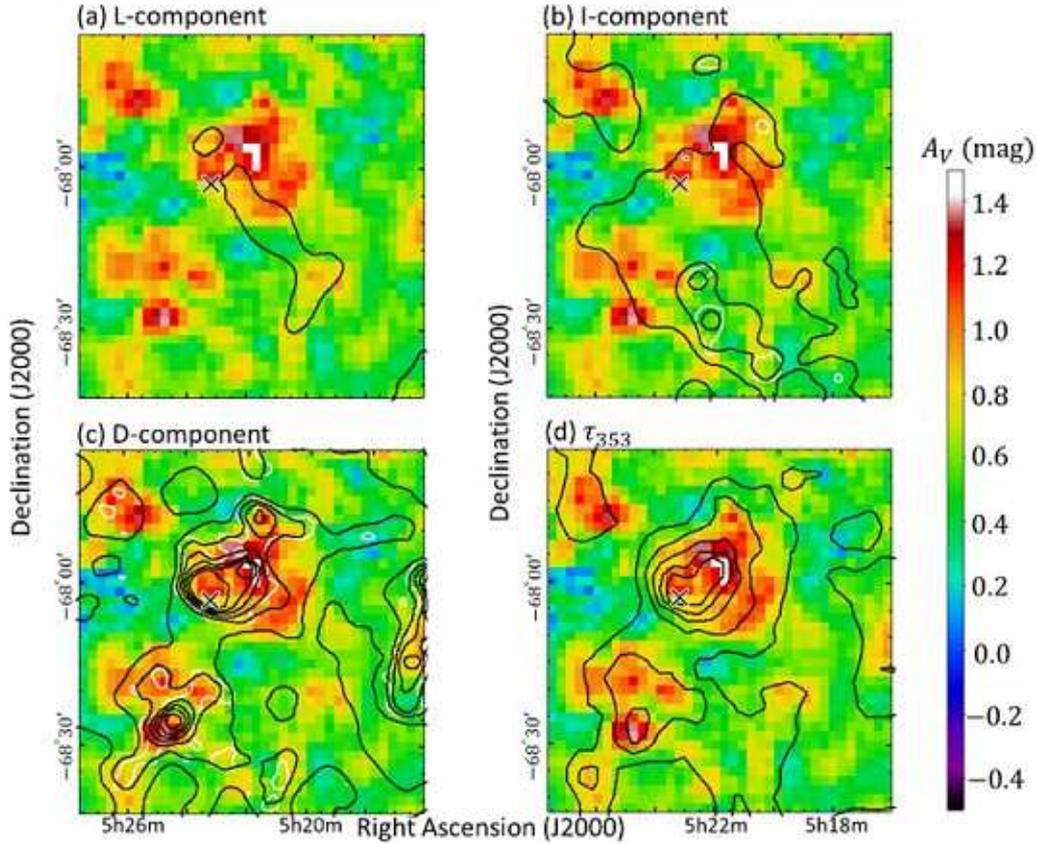} 
 \end{center}
\caption{$A_V(80\%)$ maps of N44 superposed on the $N(\rm H)$ maps of the (a) L-, (b) I- and (c) D-component shown with black contours. The contour levels are ($0.7,\ 1.6,\ 2.5,\ 3.4,\ 4.2,\ 5.1$ and $6.0$)$\times 10^{21}$ $\rm cm^{-2}$. White contours show the $1.5\sigma$ level of $^{12}\rm CO$ intensity. (d) $A_V(80\%)$ map superposed on dust emission of $\tau_{353}$ with the contour levels of ($1.6,\ 3.0,\ 4.3,\ 5.7$ and $7.0$)$\times 10^{-5}$. A cross symbol shows the position from which the plots in figure \ref{fig:eg_los} are created.}\label{fig:n44_av_w_gas}
\end{figure*}

\begin{figure*}
 \begin{center}
  \includegraphics[width=16cm]{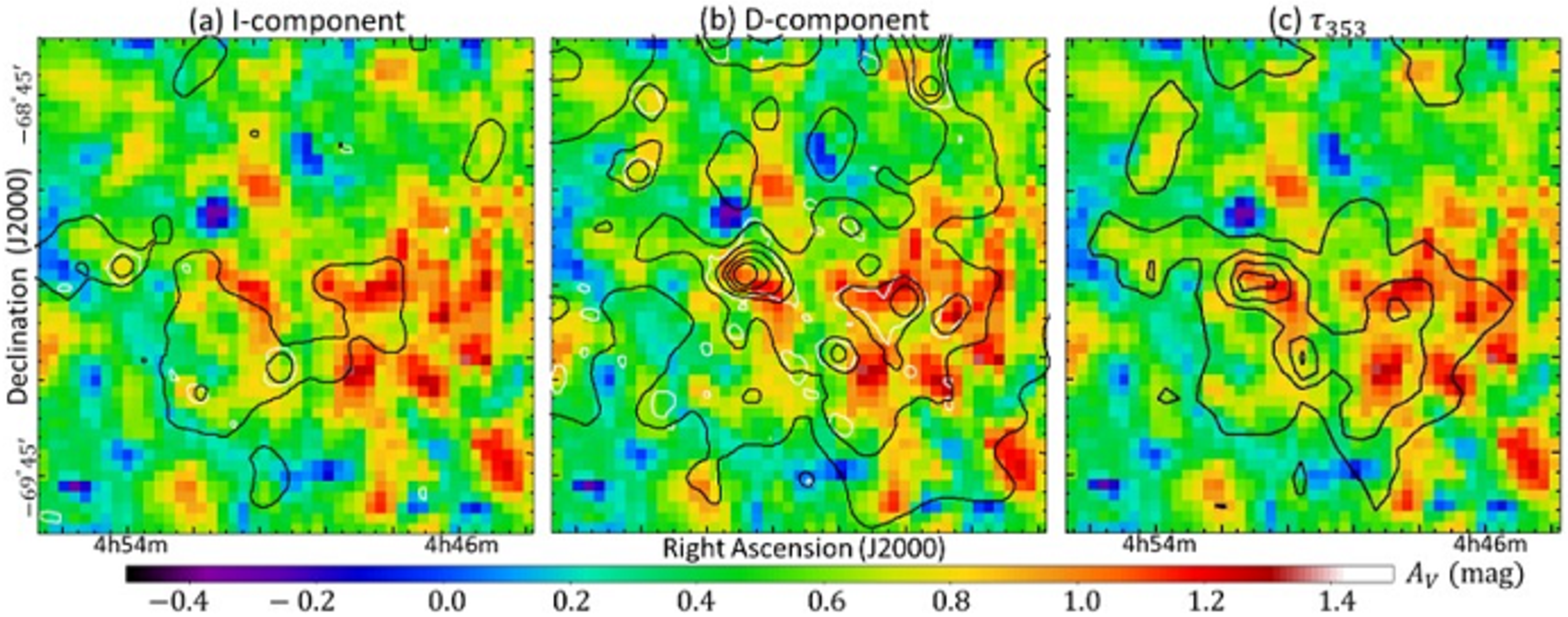} 
 \end{center}
\caption{Same maps as figure \ref{fig:n44_av_w_gas} but for N79. Contour levels are also the same as those in figure \ref{fig:n44_av_w_gas}. The comparison of $A_V(80\%)$ with $N$(H) of the L-component is not shown because the L-component is not detected significantly with these contour levels (see figure \ref{fig:av_allLMC}). The right-hand side of the map is noisy due to the low number densities of stars (figure \ref{fig:averr_allLMC}).}\label{fig:n79_av_w_gas}
\end{figure*}

\begin{figure*}
 \begin{center}
  \includegraphics[width=16cm]{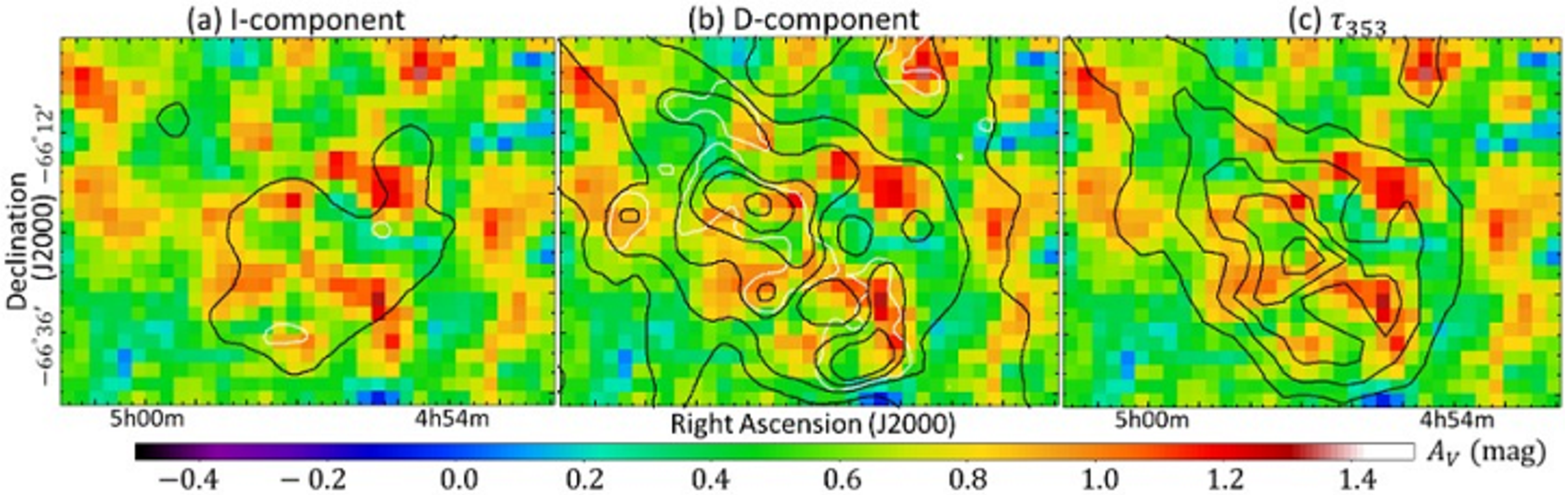} 
 \end{center}
\caption{Same maps as figure \ref{fig:n44_av_w_gas} but for N11. Contour levels are also the same as those in figure \ref{fig:n44_av_w_gas}. The comparison of $A_V(80\%)$ with $N$(H) of the L-component is not shown because the L-component is not detected significantly with these contour levels. The right-hand side of the map is noisy due to the low number densities of stars.}\label{fig:n11_av_w_gas}
\end{figure*}

In this section, to evaluate the 3D dust geometry and the dust/gas ratio of different velocity components in the massive star forming regions of N44, N79 and N11, we investigate spatial correlation between $A_V$ and $N$(H) of different velocities.
Figures \ref{fig:all_n44}--\ref{fig:percentile_n11} show the $A_V$(10\%) to $A_V$(80\%) maps of N44, N79 and N11, respectively.
Since these maps show the dust extinction integrated from observers to $X$\%, the dust extinction should increase with $X$\%.
We also present the comparison of $A_V$(80\%) with $N$(H) of different velocity components and dust emission of $\tau_{353}$ for N44, N79 and N11 in figures \ref{fig:n44_av_w_gas}--\ref{fig:n11_av_w_gas}, respectively.

The analysis area of each star forming region shown in figures \ref{fig:all_n44}--\ref{fig:percentile_n11} is determined based on the areas for which Tsuge et al. (\yearcite{tsuge}, \yearcite{tsuge_2021}) investigate the spatial and velocity distribution of the H\,\emissiontype{I} clouds.
Although they are larger than the H\, \emissiontype{II} regions of N44, N79 and N11 defined in the \citet{henize} catalog, we adopt these analysis areas covering the L- and/or I-components existing around each star forming region to investigate the 3D dust geometry of different velocity components.
\subsubsection{N44}
From figure \ref{fig:n44_av_w_gas}, we recognize that the dust extinction exhibits a good spatial correlation with $N$(H) of the D-component and $\tau_{353}$, which indicates that all the dust clouds in N44 are located in the LMC disk.
On the other hand, we cannot see a clear spatial correlation between the dust extinction and the L- and I-components.

In order to statistically determine whether the $A_V$ counterparts of the L- and/or I-components are detected in the $A_V$(80\%) map, we apply the fitting with equation (\ref{eq:regress1}) to the $A_V$(80\%) map.
Since $N$(H) of the L-component is low for a large area as shown in figure \ref{fig:n44_av_w_gas}a, to improve the statistics, we newly define the L+I-component whose velocity range is $-$100 to $-$10 $\rm km\ s^{-1}$ covering both velocity ranges of the L- and I-components, and decompose $A_V$ into the L+I- and D-components.
We first perform the fitting with the D-component alone (i.e., $k=$D in equation (\ref{eq:regress1})).
As a next step, we add the L+I-components in the fitting (i.e., $k=$D and L+I).
Here, the same $X_{\rm CO}$ factor is assumed for the D- and L+I-components (i.e., $x_{\rm D}=x_{\rm L+I}$ in equation (\ref{eq:regress1})) because the CO emission of the L- and I-components (shown by white contours in figure \ref{fig:n44_av_w_gas}) is detected in only a few areas and thus it is difficult to determine $X_{\rm CO}$ of the L+I-component by the fitting.
To judge whether the fitting is improved significantly, we perform an F-test with a confidence level of 95\% corresponding to the F-test probability of smaller than 0.05.
As a result of the fitting, the ratio of $\chi^2$ to the degree of freedom ($\chi^2$/dof) is decreased from 84.51/166 ($k=$D) to 76.67/165 ($k=$D and L+I); an F-test probability is $6\times10^{-5}$, indicating that the fitting is significantly improved by adding the L+I-component, and thus the L+I-component is significantly detected in the $A_V$(80\%) map.

To check the validity of introducing the L+I-component, we separate the L+I-component into the L- and I-components and perform the fitting with equation (\ref{eq:regress1}) using $k=$D, L and I with the same $X_{\rm CO}$ factor for the D-, L-, and I-components (i.e., $x_{\rm D}=x_{\rm L}=x_{\rm I}$).
As a result, $\chi^2$/dof is decreased from 76.67/165 ($k=$D and L+I) to 75.46/164 ($k=$D, L and I); an F-test probability is 0.11, indicating that separating the L+I-component does not improve the fitting significantly, and thus it is reasonable to introduce the L+I-component.

\begin{figure}
 \begin{center}
  \includegraphics[width=8cm]{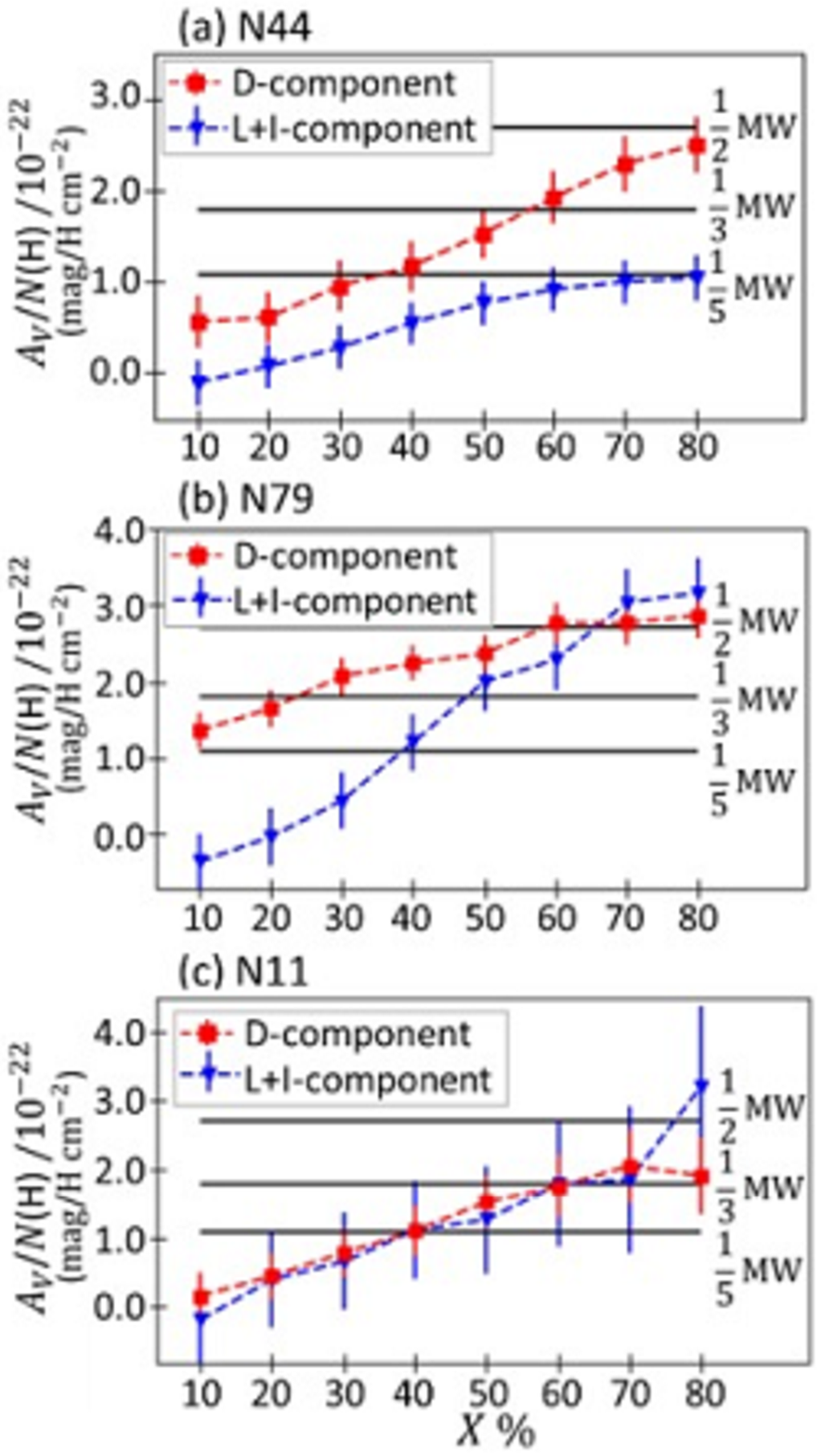} 
 \end{center}
\caption{Fitting parameters of $A_V$/$N$(H) for (a) N44, (b) N79 and (c) N11 estimated from the linear regression of equation (\ref{eq:regress1}). Squares and triangles are the parameters of the D- and L+I-components, respectively. Black horizontal lines are 1/5, 1/3 and 1/2 of $A_V$/$N$(H) in the Milky Way (MW).}\label{fig:avnh_n44}
\end{figure}

To evaluate the 3D geometry of N44, we apply the fitting with equation (\ref{eq:regress1}) using the D- and L+I-components to each $A_V$($X$\%) map for $X=$ 10\% to 80\%.
Table \ref{tab:fit_res_n44} shows the best-fit parameters, while figure \ref{fig:avnh_n44}a shows $A_V$/$N$(H) of the L+I- and D-components thus obtained as a function of $X$\%.
The uncertainties of the parameters are calculated by the formal regression errors using the $A_V$ uncertainties of $\delta A_V(X\%)$.

In table \ref{tab:fit_res_n44}, many of the derived $X_{\rm CO}$ factors of N44 for $X=$10\%--80\% are not estimated significantly (i.e., $X_{\rm CO} < 1 \sigma$, where $\sigma$ is the uncertainty of $X_{\rm CO}$).
This result indicates that $A_V$ counterparts of CO molecules are not detected in many of our $A_V$ maps.
Actually, in the total integrated $A_V$ map ($A_V$(80\%)), the $X_{\rm CO}$ factor is estimated significantly ($X_{\rm CO} > 1 \sigma$).
From table \ref{tab:fit_res_n44}, we find that $A_V/N$(H) ratio at $X=80$\% for the L+I-component of N44 is lower than that for the D-component.
Since $X_{\rm CO}$ is known to depend on the metallicty (\cite{bolatto}), we need to check the validity of using the same $X_{\rm CO}$ factor for the L+I and D-components in the fitting.
Hence, tentatively by setting $X_{\rm CO}$ for the L+I-component as a new free parameter independently from the D-component, $\chi^2$/dof is decreased from 76.67/165 ($k=$D, L+I with $x_{\rm D}=x_{\rm L+I}$) to 76.66/164 ($k=$D, L+I with $x_{\rm D} \neq x_{\rm L+I}$); an F-test probability is 0.86, which means that the fitting is not improved significantly and using the same $X_{\rm CO}$ for the L+I- and D-components does not affect our results.
This is probably due to the fact that the CO emission of the L+I-component is detected in only small areas and thus it is difficult to detect the difference in $X_{\rm CO}$ which is expected to be caused by the difference in the metallicity between the L+I- and D-components.
\begin{table*}
\tbl{Fitting parameters estimated from the comparison of $A_V$ with $N(\rm H)$ for each velocity component in the star forming regions of N44, N79 and N11.}{%
 \begin{tabular}{lccccccc}
\hline
\multicolumn{3}{c}{}&\multicolumn{2}{c}{$a_k$ $\left( \frac{A_V}{N(\rm H)} \right)$} [$10^{-22}$ mag/(H $\rm cm^{-2}$)]&&&$x_{\rm D}$ ($X_{\rm CO}$)\footnotemark[*]\\ \cline{4-5} \cline{7-8}
 Name & Percentile &Reduced $\chi^2$ & L+I-component & D-component &&& D-component\\
 \hline
N44 &$A_V$(10\%) & 0.36 &  $-0.09\pm 0.26$ &  $0.54\pm 0.30$ &&&$3.0\pm 4.6$ \\
 &$A_V$(20\%) & 0.64 &  $0.09\pm 0.25$ &  $0.62\pm 0.28$ &&&$2.5\pm 3.7$ \\
 &$A_V$(30\%) & 0.85 &  $0.33\pm 0.25$ &  $0.95\pm 0.28$ &&&$1.0\pm 2.0$ \\
 &$A_V$(40\%) & 1.01 &  $0.61\pm 0.24$ &  $1.16\pm 0.27$ &&&$0.9\pm 1.6$ \\
 &$A_V$(50\%) & 1.11 &  $0.86\pm 0.25$ &  $1.50\pm 0.28$ &&&$0.3\pm 1.2$ \\
 &$A_V$(60\%) & 1.09 &  $1.05\pm 0.26$ &  $1.88\pm 0.29$ &&&$-0.3\pm 1.0$ \\
 &$A_V$(70\%) & 0.89 &  $1.12\pm 0.26$ &  $2.26\pm 0.30$ &&&$0.2\pm 1.0$ \\
 &$A_V$(80\%) & 0.46 &  $1.11\pm 0.26$ &  $2.49\pm 0.30$ &&&$1.5\pm 1.0$ \\
N79 &$A_V$(10\%) & 0.30 &  $-0.43\pm 0.43$ &  $1.37\pm 0.25$ &&&$0.7\pm 1.9$ \\
 &$A_V$(20\%) & 0.60 &  $-0.05\pm 0.43$  &  $1.65\pm 0.25$ &&&$1.3\pm 1.6$ \\
 &$A_V$(30\%) & 0.84 &  $0.51\pm 0.45$  &  $2.06\pm 0.25$ &&&$0.2\pm 1.2$ \\
 &$A_V$(40\%) & 1.01 &  $1.38\pm 0.44$ &  $2.20\pm 0.24$ &&&$0.6\pm 1.1$ \\
 &$A_V$(50\%) & 1.22 &  $2.17\pm 0.43$  &  $2.27\pm 0.24$ &&&$1.9\pm 1.2$ \\
 &$A_V$(60\%) & 1.34 &  $2.48\pm 0.46$  &  $2.64\pm 0.25$ &&&$2.0\pm 1.1$ \\
 &$A_V$(70\%) & 1.26 &  $3.33\pm 0.50$  &  $2.62\pm 0.26$ &&&$1.9\pm 1.1$ \\
 &$A_V$(80\%) & 0.71 &  $3.30\pm 0.51$  &  $2.75\pm 0.28$ &&&$2.9\pm 1.3$ \\
N11 &$A_V$(10\%) & 0.17 &  $-0.06\pm 0.35$  &  $0.14\pm 0.31$ &&&$40.3\pm 112.6$ \\
 &$A_V$(20\%) & 0.39 &  $0.40\pm 0.63$  &  $0.45\pm 0.33$ &&&$5.1\pm 11.0$ \\
 &$A_V$(30\%) & 0.60 &  $0.65\pm 0.69$  &  $0.79\pm 0.34$ &&&$3.5\pm 6.0$ \\
 &$A_V$(40\%) & 0.73 &  $1.20\pm 0.76$  &  $1.10\pm 0.36$ &&&$0.8\pm 4.1$ \\
 &$A_V$(50\%) & 0.99 &  $1.48\pm 0.87$  &  $1.51\pm 0.38$ &&&$-0.8\pm 3.0$ \\
 &$A_V$(60\%) & 1.14 &  $2.08\pm 1.03$  &  $1.73\pm 0.46$ &&&$-1.1\pm 2.9$ \\
 &$A_V$(70\%) & 1.09 &  $2.28\pm 1.19$  &  $2.03\pm 0.54$ &&&$-0.1\pm 3.3$ \\
 &$A_V$(80\%) & 0.42 &  $3.13\pm 1.14$  &  $2.09\pm 0.53$ &&&$2.8\pm 3.3$ \\
 \hline
 \end{tabular}}\label{tab:fit_res_n44}
 \begin{tabnote}
 \footnotemark[*] The units are $10^{20}\ {\rm cm^{-2}}/(\rm{K\ km\ s^{-1}}$)\\
\end{tabnote}
 \end{table*}
\subsubsection{N79}
In figure \ref{fig:n79_av_w_gas}b, the D-component gas correlates well with the dust extinction map, although the right-hand side of the map is noisy due to the low number densities of the stars (figure \ref{fig:number_density}).
Dust emission $\tau_{353}$ also shows the spatial correlation with the dust extinction, indicating that all the dust is present in the LMC disk.

To check whether the $A_V$ counterparts of the I-component is detected significantly, we apply the same fitting procedure for N44 to the $A_V$(80\%) map of N79.
By adding the L+I-component to the fitting with the D-component alone, $\chi^2$/dof is improved from 195.28/231 ($k=$D) to 163.66/230 ($k=$D and L+I).
The F-test probability is $2\times 10^{-10}$, meaning that the fitting is significantly improved, and thus the $A_V$(80\%) map calls for the presence of the L+I-component.
The parameters estimated from the fitting with the D- and L+I-components are summarized in table \ref{tab:fit_res_n44} and plotted in figure \ref{fig:avnh_n44}b.

From table \ref{tab:fit_res_n44}, we find that $X_{\rm CO}$ for N79 is not detected significantly at $X=10\%$--40\%, while it is detected with significance higher than $1\sigma$ at $X=50\%$--80\%, which indicates that $A_V$ counterparts of CO molecules are detected only at large $X\%$.
$A_V/N$(H) at $X=80\%$ for the L+I-component of N79 is consistent with that for the D-component, and thus using the same $X_{\rm CO}$ for the L+I- and D-components in the fitting is considered to be reasonable.
\subsubsection{N11}
From figures \ref{fig:n11_av_w_gas}a and \ref{fig:n11_av_w_gas}b, we recognize that dust clouds are detected in the regions where the D- and I-components are detected.
In fact, adding the L+I-component to the fitting with the D-component alone, $\chi^2$/dof is improved from 37.07/81 to 33.93/80; an F-test probability is $8\times 10^{-3}$ and again the $A_V$(80\%) map requires the presence the L+I-component.
In figure \ref{fig:n11_av_w_gas}c, the dust extinction spatially correlates well with $\tau_{353}$, which indicates that all the dust exists in the LMC disk.
We perform the same fitting to the $A_V$($X$\%) maps of N11 as to that of N44.
The derived fitting parameters are shown in table \ref{tab:fit_res_n44} and plotted in figure \ref{fig:avnh_n44}c

From table \ref{tab:fit_res_n44}, we recognize that the $A_V/N$(H) ratio at $X=80\%$ for the L+I-component of N11 is consistent with that for the D-component within the errors, and thus using the same $X_{\rm CO}$ factor for the L+I- and D-components is considered to be reasonable.
All of $X_{\rm CO}$ for $X=10\%$--80\% is lower than $1\sigma$, which indicates that the $A_V$ counterparts of the CO molecules are not detected even in the $A_V$(80\%) map, or $X_{\rm CO}$ cannot be estimated precisely by the fitting due to the large $A_V$ uncertainties.
\subsection{Dust geometry for the Diffuse L-com region}\label{sec:diffuse_lcom}
In the ``Diffuse L-com region'' that is located in the west of N44 and is marked in figure \ref{fig:av_allLMC}, \citet{tsuge_2021} find that there is no significant H\, \emissiontype{II} region in the area where the L-component exists.
They also find that there is no signature of deceleration of the L-component in the first moment map of the Diffuse L-com region, and thus they suggest that the L-component in this region is located behind the LMC disk, and the gas collision is yet to occur.
To verify this hypothesis, we evaluate the 3D geometry of this region.

\begin{figure*}
 \begin{center}
  \includegraphics[width=14cm]{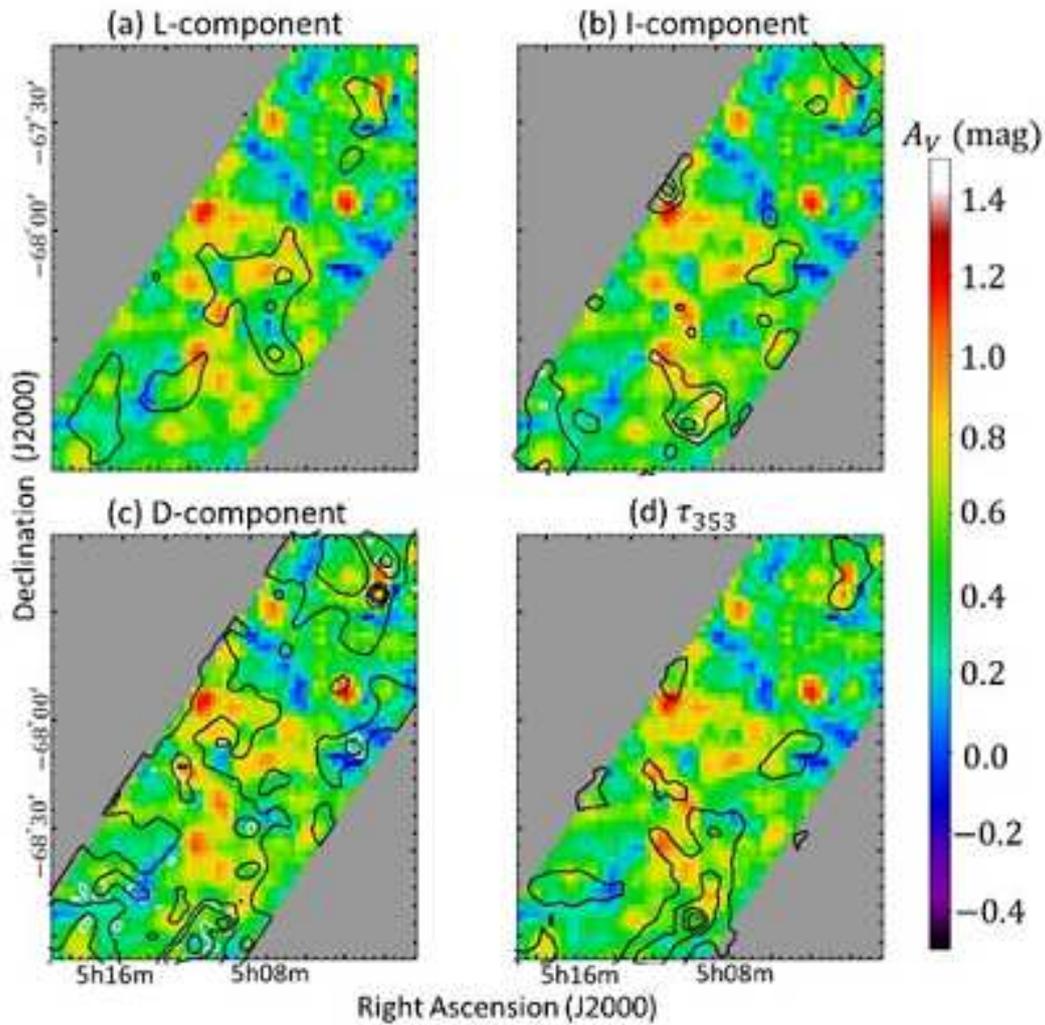} 
 \end{center}
\caption{Same maps as figure \ref{fig:n44_av_w_gas} but for the Diffuse L-com region. Grey regions are masked for the fitting with equation (\ref{eq:regress1})}\label{fig:diffuse_av_w_gas}
\end{figure*}

In figure \ref{fig:diffuse_av_w_gas}, we show the $A_V$(80\%) maps with $N$(H) of the L-, I- and D-components and $\tau_{353}$.
In figure \ref{fig:diffuse_av_w_gas}a, we cannot see a clear spatial correlation between the L-component and the dust extinction.
The D-component extends in this region and roughly correlates with the dust extinction in figure \ref{fig:diffuse_av_w_gas}c, while the CO emission of the D-component (shown by white contours in figure \ref{fig:diffuse_av_w_gas}c) does not correlate with the dust extinction especially in the south of the map, which indicates that the molecular gas is located on the far side of the LMC disk.

To determine whether the $A_V$ counterparts of the L- and/or I-component are detected significantly, we apply the linear regression fitting to the $A_V$(80\%) map similarly to N44.
Since the $A_V$ counterparts of molecular gas are not detected as mentioned above, we mask the regions in the $A_V$ map where the CO intensity is higher than 1.5$\sigma$, and perform the fitting with equation (\ref{eq:regress1}) using only the H\,\emissiontype{I} data.
By adding the L+I-component to the fitting with the D-component alone, $\chi^2$/dof is improved from 248.72/210 ($k=$D) to 230.11/209 ($k=$D and L+I); an F-test probability is $6\times 10^{-5}$, meaning that the $A_V$(80\%) map calls for the presence of the L+I-component.
In addition, as in the study of N44, we separate the L+I-component into the L- and I-components and perform the fitting using $k=$D, L and I.
As a result, $\chi^2$/dof is decreased from 230.11/209 to 223.49/208.
An F-test probability is 0.014, which means that the fitting is significantly improved by separating the L+I-component.
The estimated parameters of $A_V$/$N$(H) of the L-, I- and D-components for the $A_V$(80\%) map are ($0.37\pm0.40$, $2.63\pm0.67$ and $2.70\times0.38$)$\times 10^{-22}$ mag/$\rm cm^{-2}$, respectively.
$A_V$/$N$(H) of the L-component is not detected significantly, which indicates that the L-component is located behind the LMC disk and does not contribute to the dust extinction.
On the other hand, the I-component is significantly detected and contributes to the dust extinction.
Such a difference between the L- and I-components results in the improvement of the fitting by separating the L+I-component.

\begin{figure}
 \begin{center}
  \includegraphics[width=8cm]{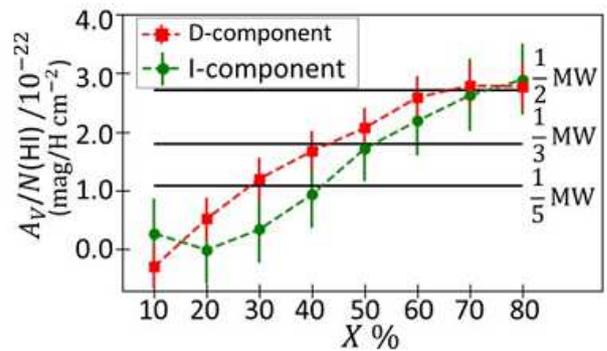} 
 \end{center}
\caption{Same diagram as figure \ref{fig:avnh_n44}, but for the Diffuse L-com region. Circles are the fitting parameters of $A_V$/$N$(H) of the I-component instead of the L+I-component shown by triangles in figure \ref{fig:avnh_n44}.}\label{fig:avnh_diffuse}
\end{figure}

To evaluate the geometry of the I- and D-components, we apply the fitting with equation (\ref{eq:regress1}) using $N$(H\,\emissiontype{I}) of the I- and D-components to each $A_V$($X$\%) map.
The estimated parameters of $A_V$/$N$(H) are summarized in table \ref{tab:fit_res_diffuse} and shown in figure \ref{fig:avnh_diffuse}.
$A_V$/$N$(H) of the D-component monotonically increases with $X$\%, indicating that the dust of the D-component extends along the line of sight.
The $A_V$/$N$(H) ratio of the I-component agrees with that of the D-component.
Thus the dust of the I-component is likely to be distributed similarly to that of the D-component.
In summary, in the Diffuse L-com region, the L-component is located behind the LMC disk, while the I- and D-components extend along the line of sight, which is consistent with the hypothesis that gas collision is yet to occur as mentioned by \citet{tsuge_2021}.
\begin{table}
\tbl{Same table as table \ref{tab:fit_res_n44} but for the Diffuse L-com region.}{%
 \begin{tabular}{lcccc}
\hline
\multicolumn{2}{c}{}&I-component&D-component\\ \cline{3-4} 
 Name &Reduced $\chi^2$ & $a_{\rm I}$ $\left( \frac{A_V}{N(\rm H)} \right)$\footnotemark[*] & $a_{\rm D}$ $\left( \frac{A_V}{N(\rm H)} \right)$\footnotemark[*] \\
 \hline
$A_V$(10\%) & 0.51 &  $0.27\pm 0.59$ &  $-0.30\pm 0.36$ \\
$A_V$(20\%) & 1.00 &  $-0.01\pm 0.59$ &  $0.52\pm 0.37$ \\
$A_V$(30\%) & 1.39 &  $0.34\pm 0.57$ &  $1.20\pm 0.35$ \\
$A_V$(40\%) & 1.65 &  $0.93\pm 0.56$ &  $1.67\pm 0.35$ \\
$A_V$(50\%) & 1.81 &  $1.71\pm 0.56$ &  $2.07\pm 0.35$ \\
$A_V$(60\%) & 1.98 &  $2.19\pm 0.59$ & $2.59\pm 0.37$  \\
$A_V$(70\%) & 1.69 &  $2.62\pm 0.62$ & $2.79\pm 0.37$  \\
$A_V$(80\%) & 1.07 &  $2.89\pm 0.61$ &  $2.77\pm 0.38$  \\
 \hline
     \end{tabular}}\label{tab:fit_res_diffuse}
 \begin{tabnote}
 \footnotemark[*] The units are $10^{-22}$ mag/(H $\rm cm^{-2}$) \\
\end{tabnote}
 \end{table}

\section{Discussion}
\subsection{Dust geometry of individual star forming regions}\label{dis:geometry}
\begin{figure*}
 \begin{center}
  \includegraphics[width=16cm]{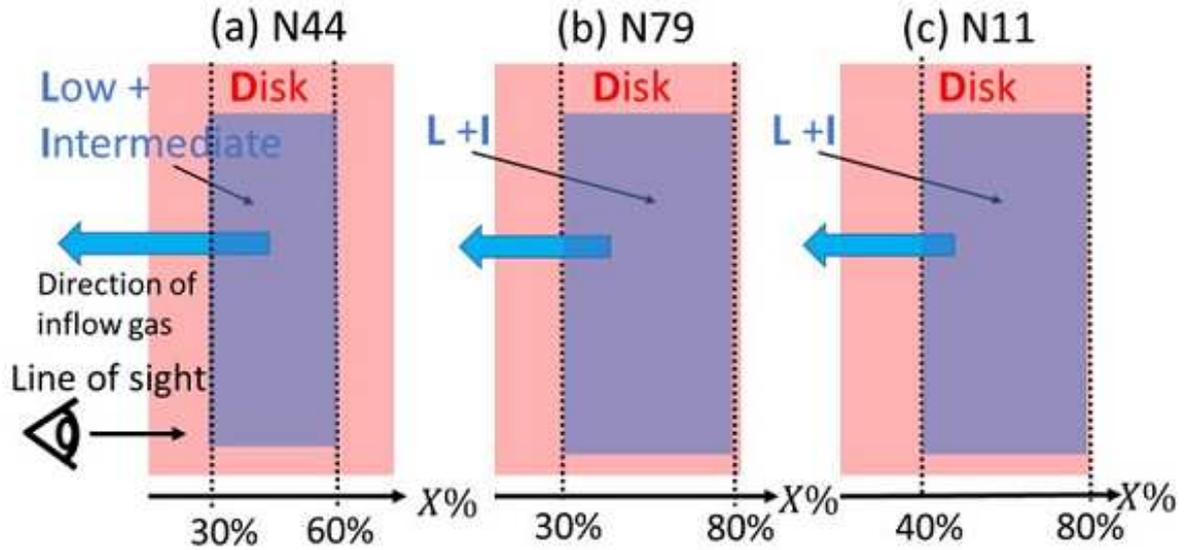} 
 \end{center}
\caption{Side view of the expected dust geometry of (a) N44, (b) N79 and (c) N11. Red and blue rectangles denote the disk velocity component (D-component) and the low+intermediate velocity component (L+I-component), respectively.}\label{fig:geometry}
\end{figure*}

In this section, we evaluate the dust geometry of N44, N79 and N11 from the estimated fitting parameter of $a_k$ (i.e., $A_V$/$N$(H)) of each $X$\%.
We first consider the dust geometry of N44.
The likely 3D dust geometry of the D- and L+I-components for N44 is illustrated in figure \ref{fig:geometry}a.
First, $A_V$/$N$(H) of the D-component in figure \ref{fig:avnh_n44}a monotonically increases with $X$\%, and thus the D-component is expected to extend along the line of sight as shown by a red rectangle in figure \ref{fig:geometry}a, which is consistent with a general idea that the gas of the LMC disk is mixed with stars.
$A_V$/$N$(H) of the L+I-component is significantly detected at $X=$30\% and onward, which indicates that the head of the L+I-component is located at $X$=30\%.
In addition, $A_V$/$N$(H) of the L+I-component is almost constant from $X$=60\% to 80\%, indicating that there is little dust of the L+I-component beyond $X=60$\%.
As a whole, the L+I-component in N44 is likely to extend from $X=30$\% to 60\% as shown by a blue rectangle in figure \ref{fig:geometry}a.
The dust geometry suggests that the L+I-component is penetrating the D-component.
This trend supports that gas collision between the gas of the LMC disk and an inflow gas as seen in the L+I-component may have induced the massive star formation in N44 (\cite{tsuge}).

Similarly to N44, we discuss the 3D dust geometry of N79 and N11.
In both regions, from the monotonically increase of $A_V$/$N$(H) of the D-component with $X$\% in figures \ref{fig:avnh_n44}b and \ref{fig:avnh_n44}c, the D-component is expected to extend along the line of sight as shown by red rectangles in figures \ref{fig:geometry}b and \ref{fig:geometry}c.
$A_V$/$N$(H) of the L+I-component is not detected significantly at $X=$10\%--20\% and $X=$10\%--30\% for N79 and N11, respectively (see table \ref{tab:fit_res_n44}), and thus the L+I-component is expected to extend beyond $X=30$\% for N79 and beyond $X=40$\% for N11 as shown by blue rectangles in figures \ref{fig:geometry}b and \ref{fig:geometry}c, respectively.

We compare the dust geometry of N44, N79 and N11 as well as the H\,\emissiontype{I} ridge region with the evolutionary stages of giant molecular clouds (GMCs) proposed by \citet{kawamura}.
We summarize the positions of the L- or L+I-components and evolutionary stages of the relevant GMCs in table \ref{tab:summary}.
The H\,\emissiontype{I} ridge region is separated into three regions based on the geometry of the L-component by \citet{furuta_2021}.
They define them as regions 1--3 from north to south; regions 1 and 2 contains 30 Dor and N159, respectively (see figure 14 in \cite{furuta_2021}).
The evolutionary stages are classified into four stages, Type $\rm{\,I\,}$--$\rm{I\hspace{-.15em}I\hspace{-.15em}I}$ and the last stage in order from youngest to oldest.
From the positions relative to the LMC disk of the L+I-component for N44, N79 and N11 in table \ref{tab:summary}, we recognize that the gas collision in these regions occurred later than in 30 Dor but occurred earlier than in region 3, which is consistent with the evolutionary sequences of the GMCs.
The evolutionary stage of GMCs in N79 is earlier than those of N159, N44 and N11.
However, we cannot see clear differences in the timing of gas collision between these regions.
This is likely to be caused by the large $A_V$ uncertainties especially for N79 and N11 due to the low number densities of the stars.

In order to discuss the relationship between the timing of the gas collision and the evolutionary stages of GMCs quantitatively, we compare the crossing time scale of the gas collision with the transition time scale of the evolutionary stages.
For example, from the comparison of the dust geometry of 30 Dor with that of N44, the head of the L+I-component is separated by the distance of 30\% percentile.
Considering the LMC disk thickness of 2 kpc (\cite{thickness}), this corresponds to 0.6 kpc.
Assuming the velocity of the inflow gas (L-component) $\sim 100$ km/s (\cite{tsuge_2021}), it takes 6 Myr for the inflow gas to cross 0.6 kpc, which is consistent with the time scale of 7 Myr for the transition from Type $\rm{I\hspace{-.15em}I\hspace{-.15em}I}$ GMCs in N44 to the last-stage GMCs in 30 Dor (\cite{kawamura}).
Therefore, our result can reasonably explain the difference of the evolutionary stages of the GMCs across the LMC.
\begin{table*}
\tbl{Comparison of dust geometry of the L- or L+I-component with evolutionary stages of the related giant molecular clouds.}{%
 \begin{tabular}{lccc}
\hline
 Name & Position & Evolutionary stage\footnotemark[*] & \\
  & (L- or L+I-component) & \\
\hline
30 Dor (region 1 in H\,\emissiontype{I} ridge) & front of the disk\footnotemark[$\dagger$] & last-stage \\
N159 (region 2 in H\,\emissiontype{I} ridge) & $X=30$\% to 80\%\footnotemark[$\dagger$] & Type $\rm{I\hspace{-.15em}I\hspace{-.15em}I}$ \\
region 3 in H\,\emissiontype{I} ridge& behind the disk\footnotemark[$\dagger$] & Type $\rm{\,I\,}$ \\
N44 & $X=30$\% to 60\% & Type $\rm{I\hspace{-.15em}I\hspace{-.15em}I}$ \\
N79 & $X=30$\% to 80\% & Type $\rm{I\hspace{-.15em}I}$ \\
N11 & $X=40$\% to 80\% & Type $\rm{I\hspace{-.15em}I\hspace{-.15em}I}$ \\
\hline
 \end{tabular}}\label{tab:summary}
 \begin{tabnote}
 \footnotemark[*] \citet{kawamura} \\
 \footnotemark[$\dagger$] \citet{furuta_2021} \\
\end{tabnote}
 \end{table*}
\subsection{Origin of massive star forming region}\label{sec:dis_avnh}
\begin{table*}
\tbl{$A_V$/$N$(H) and $X_{\rm CO}$ at $X=80\%$ of each velocity component for the H\,\emissiontype{I} ridge region, N44, N79 and N11.}{%
 \begin{tabular}{lccccccc}
\hline
 &L or L+I-component&I-component& &\multicolumn{2}{c}{D-component}\\ \cline{2-3} \cline{5-6} 
 Name & $a_{\rm L}$ $\left( \frac{A_V}{N(\rm H)} \right)$\footnotemark[*] & $a_{\rm I}$ $\left( \frac{A_V}{N(\rm H)} \right)$\footnotemark[*]&& $a_{\rm D}$ $\left( \frac{A_V}{N(\rm H)} \right)$\footnotemark[*]&$x_{\rm D}$ $\left(X_{\rm CO}\right)$\footnotemark[$\dagger$] \\
 \hline
H\,\emissiontype{I} ridge & $1.24\pm 0.13$\footnotemark[$\ddagger$] & $1.36\pm 0.17$\footnotemark[$\ddagger$] &&$2.08\pm 0.14$\footnotemark[$\ddagger$] & $1.7\pm 1.0$\footnotemark[$\ddagger$]\\
N44 & $1.11\pm 0.26$ & - && $2.49\pm 0.30$ &$1.5\pm 1.0$ \\
N79 & $3.30\pm 0.51$ & - && $2.75\pm 0.28$ &$2.9\pm 1.3$ \\
N11 & $3.13\pm 1.14$ & - && $2.09\pm 0.53$ &$2.8\pm 3.3$ \\
 \hline
     \end{tabular}}\label{tab:fit_res_all}
 \begin{tabnote}
 \footnotemark[*] The units are $10^{-22}$ mag/(H $\rm cm^{-2}$) \\
 \footnotemark[$\dagger$] The units are $10^{20}\ {\rm cm^{-2}}/(\rm{K\ km\ s^{-1}}$)\\
 \footnotemark[$\ddagger$] \citet{furuta_2021}\\
\end{tabnote}
 \end{table*}
We discuss the origins of the massive star forming regions of N44, N79 and N11 from the $A_V$/$N$(H) ratio at $X$=80\%.
We summarize $A_V$/$N$(H) for these regions as well as the H\,\emissiontype{I} ridge region in table \ref{tab:fit_res_all}.
\citet{furuta_2021} find the difference by a factor of two in $A_V$/$N$(H) between the D- and the other components for the H\,\emissiontype{I} ridge region (see table \ref{tab:fit_res_all}).
From the existence of the low metallicity gas, they propose the massive star formation in 30 Dor triggered by galactic interaction between the LMC and the SMC.

The $A_V$/$N$(H) ratio of the D-component at $X=$80\% for the star forming regions of N44, N79 and N11 is similar to that of the H\,\emissiontype{I} ridge region (see table \ref{tab:fit_res_all}).
The resultant $A_V$/$N$(H) ratio is nearly a half of that in the Milky Way ($5.34\times10^{-22}$ mag $\rm cm^{2}$ for $R_V$=3.1; \cite{galactic_dustgas}), which is consistent with the dust/gas ratio for the LMC of $\sim$1/2--1/3 Galactic value (\cite{gdr_smc_spitzer}; \cite{gdr_herschel}).
The $X_{\rm CO}$ factors of the D-component at $X=80\%$ are consistent within the errors between all the star forming regions shown in table \ref{tab:fit_res_all}, although the uncertainties are large especially in N11.
The resultant $X_{\rm CO}$ factors for all the star forming regions are similar to that of the LMC estimated from CO and [CII] observations after correcting the photodisociation of CO molecules due to UV radiation ($X_{\rm CO} \sim 2.9\times10^{20}$ ${\rm cm^{-2}}/(\rm{K\ km\ s^{-1}})$; \cite{corr_xco}).

On the other hand, $A_V$/$N$(H) of the L+I-component significantly varies between the star forming regions.
For N44, $A_V$/$N$(H) of the L+I-component is $\sim$1/5 of the Galactic value, which is similar to the dust/gas ratio for the SMC of $\sim$1/6 Galactic value (\cite{gdr_herschel}).
The existence of the low metallicity gas of the L+I-component in N44 supports the contamination of the inflow gas from the SMC as proposed by \citet{tsuge}.

In N44, expansion of H\,\emissiontype{I} shell is reported (\cite{kim}).
Around the shell, three episodes of massive star formation are found; one is the $\sim10$ Myr-old star formation inside the shell, another is the $\sim5$ Myr-old star formation on the shell rims, and the other is the YSOs (typically $<$ 1 Myr-old) (e.g., \cite{oey}; \cite{chen09}; \cite{car12}).
\citet{tsuge} propose that the 5 Myr-old massive star formation was triggered by the galactic interaction between the LMC and the SMC as an alternative scenario of star formation triggered by the expansion of the H\,\emissiontype{I} shell from the existence of low metallicity gas and signature of gas collision between the L- and I-components in addition to the lack of energy supply from the shell to explain the motions of the L-, I- and D-components in N44.
The age of the stellar population of N44 ($\sim5$ Myr) is younger than that around the 30 Dor region of $8.1$ Myr (\cite{sch18}), which agrees with the difference in the evolutionary stages of GMCs between N44 and 30 Dor (see table \ref{tab:summary}).
Therefore, from the low $A_V$/$N$(H) ratio of the L-component and the gas colliding geometry in figure \ref{fig:geometry}a, we suggest that the massive star formation  around the shell rims in N44 was triggered by the galactic interaction between the Magellanic Clouds, similarly to the H\,\emissiontype{I} ridge region.

For N79 and N11, $A_V$/$N$(H) of the L+I-component at $X$=80\% is nearly a half of the Galactic value, which is similar to that of the D-component.
According to the recent numerical simulations of the tidal interaction between the LMC and SMC (\cite{tsuge_2021}),
the gas currently falling onto the LMC disk as the L+I-component consists of the gas from the LMC as well as from the SMC. 
We thus expect in some places that the metallicity is not much different from that of the LMC, if the L+I-component is dominated locally by the LMC gas. 
In addition, an analysis of the dust/gas ratio based on the Planck dust emission of $\tau_{353}$ (Tsuge et al. 2021b, in preparation) indicates that the dust/gas ratio toward N79 and N11 is about two times larger than that in the H\,\emissiontype{I} ridge region and N44.  
This is consistent with the present results. 
It is therefore possible that the N79 and N11 regions were triggered by the L+I-component which was originated in the LMC. 
We thus suggest that the trigger in the two regions was due to the internal interaction although the possibility of the tidal interaction cannot be ruled out because the numerical simulation by \citet{tsuge_2021} suggests that part of the LMC gas is stripped off from the LMC disk in the interaction and then merge with the SMC gas to fall down to the LMC disk.

N79 is located at the intersection of the galactic bar-end with the spiral arm.
\citet{n79_nat} suggest that the unique location may provide the gas accumulation and compression to create massive star formations.
A similar case is reported for W43 which is located at the intersection of the Galactic bar-end with the Scutum Arm.
\citet{kohno} find several velocity components with a velocity difference of $\sim$20 km $\rm s^{-1}$ in W43 and suggest that gas collision between them triggered the starburst.
The velocity difference is similar to that between the D- and I-components in our study.
Therefore, we suggest that the massive star formation in N79 was triggered by collision of the gas in the stellar-bar with inflow gas from the spiral arm. 

In N11, several OB associations are located at the periphery of a central cavity with a diameter of 170 pc evacuated by the rich OB association of LH9 (\cite{lucke}).
Previous studies suggest that the OB association of LH10 located in the peripheral clouds was triggered by the expansion of the supershell blown by LH9 (e.g., \cite{barba}; \cite{hatano}; \cite{celis}).
The expansion velocity of the supershell is estimated to be $\sim 20\ \rm km\  s^{-1}$ (\cite{meaburn}), which agrees with the difference in the velocity between the I- and D-components.
The spatial distribution of the I-component is similar to the supershell structure (\cite{tsuge_2021}).
Using the I-component velocity of 30 km/s, it takes 2.8 Myr for the I-component to travel the radius of the supershell (85 pc), which is similar to the difference in the age of 2 Myr between LH9 and LH10 (\cite{wal92}).
Therefore, our results support that the massive star formation of LH10 in N11 was triggered by internal interaction between the supershell and the surrounding ISM.

In summary, when $N$(H) of the L-component is detected significantly as in the cases of the H\,\emissiontype{I} ridge region and N44, $A_V$/$N$(H) of the L+I-component is lower than that of the D-component.
It is likely that the L+I-component originates from inflow gas from the SMC and the star formation is triggered by external interaction between the Magellanic Clouds.
On the other hand, in the cases that $N$(H) of the L-component is not detected significantly as in the cases of N79 and N11, $A_V$/$N$(H) of the L+I-component shows fair agreement with that of the D-component.
In these regions, internal interactions such as gas converging from the spiral arm and the expansion of a supershell are likely to trigger the massive star formation.
\section{Conclusion}
We derive the three-dimensional dust extinction map of the massive star forming regions of N44, N79 and N11 in the entire LMC using the percentile method proposed by \citet{furuta_2021}.
Our total integrated dust extinction map calls for more abundant dust in the star forming regions than the previous near-infrared dust extinction map constructed by \citet{dobashi}.
From the comparison of the dust extinction with the hydrogen column densities of different velocity components, we investigate the three-dimensional geometry and the dust/gas ratio of the different velocity components for each star forming region.
Our main results are as follows:
\begin{enumerate}
\item The dust geometry of N44, N79 and N11 suggests that the L+I-component is penetrating the LMC disk.
Considering the present results together with the dust geometry of the H\,\emissiontype{I} ridge region estimated by \citet{furuta_2021}, the difference in the timing of the gas collision agrees with the difference in the evolutionary stages of the giant molecular cloud related to each star forming region (\cite{kawamura}).

\item In the Diffuse L-com region, the dust geometry indicates that the L-component is located behind the LMC disk.
This geometry is consistent with the hypothesis that the gas collision is yet to occur in this region, as expected from no signature of deceleration of the L-component and no existence of massive stars (\cite{tsuge_2021}).

\item The $A_V$/$N$(H) values and the $X_{\rm CO}$ factors of the D-component at $X=80\%$ are almost the same for all the star forming regions, while the $A_V/N$(H) ratios of the L+I-component vary between the star forming regions.
For N44, $A_V$/$N$(H) of the L+I-component is 1/5 of the Galactic value, which is similar to the dust/gas ratio for the SMC (\cite{gdr_herschel}).
From the low  $A_V$/$N$(H) ratio, we suggest that external interaction between the LMC and the SMC triggered the star formation in N44, similarly to 30 Dor in the H\,\emissiontype{I} ridge region.

\item For N79 and N11, the $A_V$/$N$(H) ratio of the L+I-component is 1/2 of the Galactic value, which is similar to that of the D-component.
Thus, we suggest the star formations in these regions were triggered by internal interaction although the possibility of the tidal interaction cannot be ruled out.
We suggest that the massive star formation of N79 was triggered by a converging gas flow from the spiral arm based on the unique location of N79, while that of N11 was triggered by interaction between the expansion of a supershell and the surrounding ISM.
\end{enumerate}

\begin{ack}
We thank Prof. Kazuhito Dobashi for kindly giving us the data of their $A_V$ map.
We also thank the referee for giving us helpful comments.
The IRSF project is a collaboration between Nagoya University and the SAAO supported by the Grants-in-Aid for Scientific Research on Priority Areas (A) (Nos. 10147207 and 10147214) and Optical \& Near-Infrared Astronomy Inter-University Cooperation Program, from the Ministry of Education, Culture, Sports, Science and Technology (MEXT) of Japan and the National Research Foundation (NRF) of South Africa.
This research was financially supported by Grant-in-Aid for JSPS Fellows Grant Number 20J12119.
\end{ack}

%

\end{document}